\documentclass[10pt,journal]{IEEEtran}
\usepackage{balance}
\usepackage{amssymb}
\usepackage{stmaryrd}
\usepackage{cite}
\usepackage{color}
\usepackage{multirow}
\usepackage{graphicx,times}
\usepackage{epstopdf}
\usepackage{indentfirst}
\usepackage{CJK}
\usepackage{amsmath}
\usepackage{amsfonts}
\usepackage{txfonts}
\usepackage{mathrsfs}
\usepackage{subfigure}
\usepackage{graphicx}
\usepackage{theorem}
\usepackage{algorithm}
\usepackage{algorithmic}
\usepackage{url}

\newtheorem{proposition}{Proposition}

\newtheorem{corollary}{Corollary}

\newtheorem{remark}{Remark}

\begin{document}
\title{Authenticating On-Body IoT Devices: An Adversarial Learning Approach}
\author{\IEEEauthorblockN{Yong Huang, Wei Wang,~\IEEEmembership{Senior Member,~IEEE}, Hao Wang, Tao Jiang,~\IEEEmembership{Fellow,~IEEE}, Qian Zhang,~\IEEEmembership{Fellow,~IEEE}}
	\thanks{Part of this work has been presented at IEEE ICC \cite{Yong2019onbody}.}
	\thanks{This work was supported in part by the National Key R\&D Program of China under Grant 2017YFE0121500, 2019YFB180003400, Young Elite Scientists Sponsorship Program by CAST under Grant 2018QNRC001, National Science Foundation of China with Grant 91738202, RGC under Contract CERG 16203719, 16204418.\textit{ (Corresponding author: Wei Wang.)  }}
	\thanks{Y. Huang, W. Wang and T. Jiang are with the School of Electronic Information and Communications, Huazhong University of Science and Technology, Wuhan 430074, China (e-mail:\{yonghuang, weiwangw, taojiang\}@hust.edu.cn).}
	\thanks{H. Wang is with Computer Science and Artificial Intelligence Lab, Massachusetts Institute of Technology, Cambridge, MA 02139 USA (e-mail:hwang87@mit.edu).}
	\thanks{Q. Zhang is with Department of Computer Science and Engineering, Hong Kong University of Science and Technology, Clear Water Bay, Hong Kong, China (e-mail:qianzh@cse.ust.hk).}}	

\maketitle

\begin{abstract}	
By adding users as a new dimension to connectivity, on-body Internet-of-Things (IoT) devices have gained considerable momentum in recent years, while raising serious privacy and safety issues. Existing approaches to authenticate these devices limit themselves to dedicated sensors or specified user motions, undermining their widespread acceptance. This paper overcomes these limitations with a general authentication solution by integrating wireless physical layer (PHY) signatures with upper-layer protocols. The key enabling techniques are constructing representative radio propagation profiles from received signals, and developing an adversarial multi-player neural network to accurately recognize underlying radio propagation patterns and facilitate on-body device authentication. Once hearing a suspicious transmission, our system triggers a PHY-based challenge-response protocol to defend in depth against active attacks. We prove that at equilibrium, our adversarial model can extract all information about propagation patterns and eliminate any irrelevant information caused by motion variances and environment changes. We build a prototype of our system using Universal Software Radio Peripheral (USRP) devices and conduct extensive experiments with various static and dynamic body motions in typical indoor and outdoor environments. The experimental results show that our system achieves an average authentication accuracy of 91.6\%, with a high area under the receiver operating characteristic curve (AUROC) of 0.96 and a better generalization performance compared with the conventional non-adversarial approach. 
\end{abstract}


\maketitle

\section{Introduction}
Human-centric Internet of Things (IoT) has recently gained increasing popularity in both industrial and academic fields by adding users as a new dimension to connectivity and enabling intriguing user-centered applications, such as remote healthcare and real-time activity tracking \cite{chiang2016fog,lai2019vifi}.  The minimalist design paradigm of IoT devices appears to be two sides of the same coin: it allows ultra-low-power communications while rendering communication links vulnerable to malevolent attackers. Since on-body IoT devices are generally attached to users' bodies to continuously record fine-grained vital signs, security breaches of these devices pose a serious threat to users' everyday privacy \cite{wang2015privacy,wang2016} and safety \cite{gollakota2011they,Luo2018authenticating}.

Despite growing attempts and extensive endeavors, it is still challenging to thwart invaders for hardware-constrained on-body IoT devices \cite{wang2019corss,shuai2019networking,kim2017free}. Recent efforts have demonstrated the feasibility of exploiting radio characteristics in body area networks (BANs) to facilitate device verification \cite{shi2013bana,Li2018secret}. Moreover, dedicated sensors, such as accelerometers \cite{revadigar2017accelerometer} and gyroscopes \cite{xu2017gait}, have been leveraged to authenticate wearable devices. However, hardly any of them have obtained prevalent adoption. They either require the assistance of specialized user motions \cite{shi2013bana,Li2018secret}, or are confined to fitness-related wearables \cite{revadigar2017accelerometer,xu2017gait}. To embrace the coming wave of human-centric IoT, it is critical for a device authentication solution to support various on-body IoT devices without specified user motions in diverse environments.

\begin{figure}
	\centering
	\includegraphics[width=0.95\linewidth]{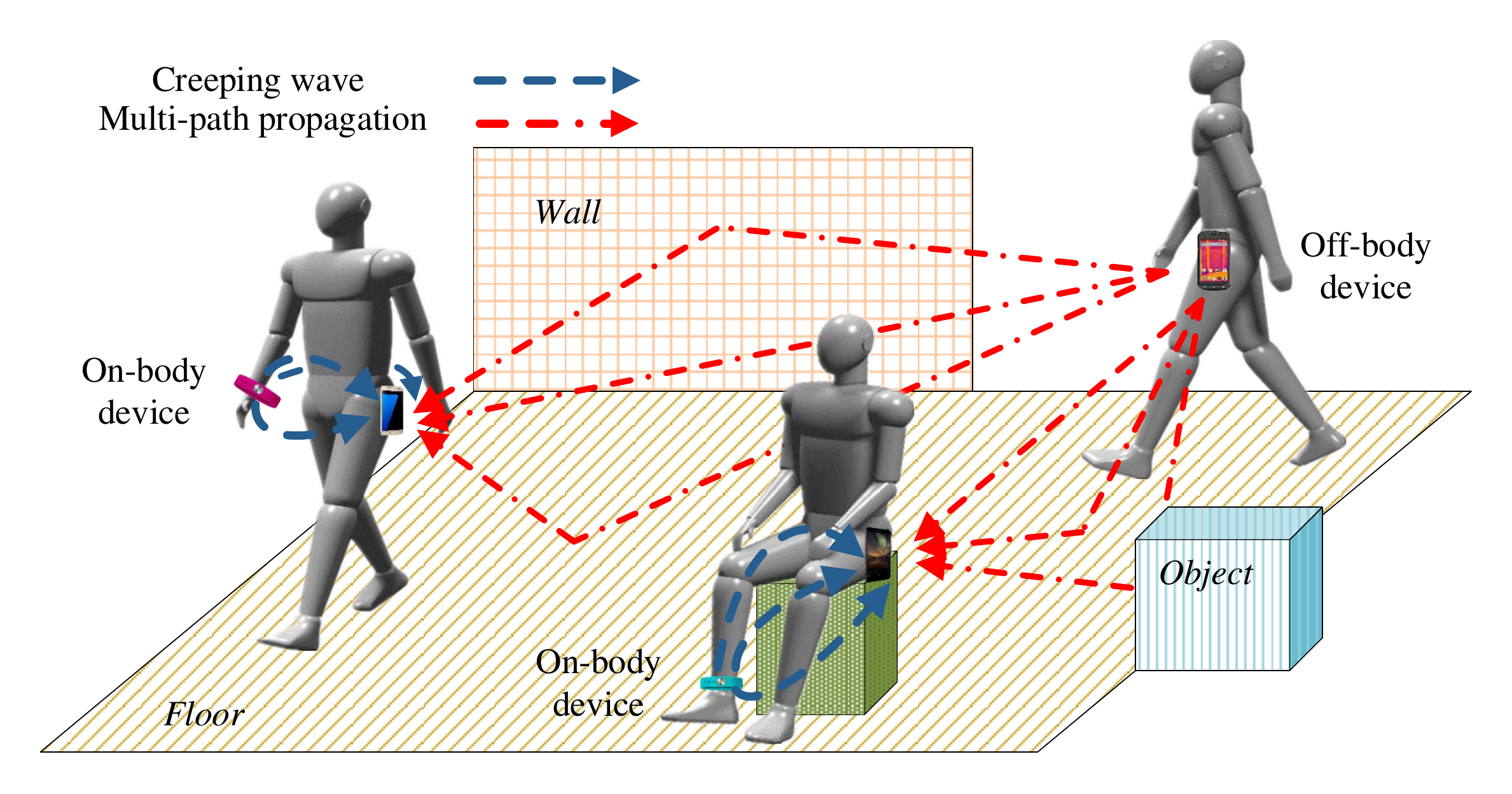}
	\caption{On- and off-body radio propagations. On-body signals are dominated by creeping waves, while off-body signals are mainly comprised of LOS and multi-path components.}
	\label{fig:propagation-patterns}
\end{figure} 

The salient physical layer (PHY) signatures naturally underlying different BANs present us with an exciting opportunity. As depicted in Fig.~\ref{fig:propagation-patterns}, for off-body wireless links, where a transmitter (Tx) and a receiver (Rx) are placed on different human bodies, radio signals are mainly comprised of direct line-of-sight (LOS) and multi-path components. On the other hand, for on-body links, where a Tx-Rx pair is carried on the same body, radio signals are governed by \textit{creeping waves} \cite{di2011body,ryckaert2004channel,kim2017exploiting,wang2015corLayer}. The distinct radio propagation patterns potentially enable a general security solution relying on prevalent wireless chips. However, radio signals in BANs are severely affected by IoT users' body motions and surrounding environments. As a consequence, on- and off-body signals can exhibit significantly different patterns under a specific user motion in a specific ambient environment, and their patterns tend to vary dramatically under a different motion in a new environment. Furthermore, the frequent change of users' motion and location in daily life makes it a highly challenging task to manually select features to represent propagation patterns from real-world radio traces. 
 
To address this challenge, we propose a motion and environment invariant authentication framework for on-body IoT devices by exploiting distinct BAN radio propagation signatures. The basic ideas lying in the proposed system are effectively constructing representative radio propagation profiles from received signals, and leveraging a neural network to essentially recognize propagation patterns and thus verify on-body IoT devices anytime and anywhere. 

We realize the above ideas by answering the following two questions.

\textit{1) How to obtain effective information on radio propagation patterns from received signals?} The received radio signals from real-world environments typically comprise massive noisy components due to complex environmental dynamics and unwanted radio interference, which makes it unlikely to recognize radio propagation patterns directly from such noisy signals. Therefore, it is crucial to extract fine-grained radio features from raw signals. In our experiments, we observe that distinctive radio propagation signatures can be represented in the time and frequency domains of received signal strength (RSS) segments. Based on this observation, we construct effective radio propagation profiles that contain representative time and frequency domain features from RSS segments for subsequent propagation pattern recognition.

\textit{2) How to learn a neural network that generalizes well in unseen scenarios?} Radio features extracted from RSS segments generally convey substantial information that is specific to ongoing user motions and surrounding environments. As a result, a neural network that is trained under a specific motion in a specific environment will undoubtedly not work well when being applied to verify devices under another motion in a different environment. To overcome this predicament, we develop an adversarial multi-player network for robust device authentication. Particularly, our network includes four functional components: a \textit{Feature Extractor}, an \textit{On-Off Predictor}, a \textit{Domain Discriminator} and an \textit{Environment Classifier}. To learn transferable features, we implement an adversarial training criterion, where the predictor works together with the extractor to learn radio propagation patterns, and both the discriminator and classifier, meanwhile, force the extractor to selectively eliminate motion and environment specific features from itself. After this training process, the extractor and predictor are expected to be resilient to unseen user motions and environments.

\textbf{Summary of results.} We implement a working prototype of our authentication system with three Universal Software Radio Peripheral (USRP) devices and conduct extensive experiments under frequently appearing body motions in multiple indoor and outdoor environments. The experimental results show that our system achieves an accuracy of 91.6\%, with an area under the receiver operating characteristic curve (AUROC) of 0.96. Specifically, it can successfully recognize 90.6\% of legitimate devices and at the same time mitigate 92.8\% of active attacks.

\textbf{Contributions.} The main contributions of this work are summarized as follows.
\begin{itemize}
	\item We propose a general authentication system that secures various on-body IoT devices without specified user motions in diverse environments. The crux of the proposed system is to construct reliable radio propagation profiles from RSS segments and to develop an adversarial multi-player neural network for essentially identifying on-body IoT devices.
    \item We theoretically analyze our adversarial network and prove that at equilibrium, the learned feature representation contains all information about BAN radio propagation patterns, and becomes invariant to motion variances and environment changes.
    \item We build a prototype of our system on USRP devices and conduct extensive experiments with various frequently appearing body motions in a variety of indoor and outdoor environments. The experimental results demonstrate the effectiveness and generalizability of our system.
\end{itemize} 

The remainder of this paper is organized as follows. The literature review is provided in Section~\ref{sec: relatedwork}. In Section~\ref{sec: motivation}, we illustrate the distinct radio signatures in different BAN channels. In Section~\ref{sec: overview}, we sketch the main design of our device authentication system and its integration with upper-layer protocols. Next, Section~\ref{sec: profile} details the construction of radio propagation profiles. Then, Section~\ref{sec: recognition} elaborates on our adversarial multi-player network for verifying on-body IoT devices. Section~\ref{sec: evaluation} shows the evaluation results. Finally, the paper is concluded in Section~\ref{sec: conclusion}.

\section{Related Work}\label{sec: relatedwork}

\textbf{Device/user authentication.} Spurred by the prevalence of wearable devices, user/device authentication has already drawn significant interest in the academic community \cite{Tomasin2018}. Dedicated sensors, including accelerometers \cite{revadigar2017accelerometer}, biometric \cite{li2015Bio} and acoustic sensors \cite{Halevi2013Acoustic}, are widely used to infer identities of wearable devices. Moreover, motion sensors \cite{xu2016walkie} are also leveraged to check if wearable devices share unique movement patterns when device carriers are in the walking state. However, sensor-based approaches limit themselves to fitness-related wearables or to sports scenarios. In contrast, our system takes advantage of pervasive wireless chips embedded in IoT devices and enables device verification under static and dynamic user body motions. 

Besides assistance from auxiliary sensors, underlying PHY signatures in BANs are also examined for verifying wearable devices. There have been many studies on the channel measurements of BANs \cite{di2011body,ryckaert2004channel,alves2011analytical,hu2007measurements}, which reveal essential differences between on- and off-body radio propagations. RSS variances are calculated to identify wearable devices in healthcare applications\cite{shi2013bana}. Furthermore, creeping waves are also exploited in \cite{wang2017securing} to secure on-body devices, wherein small- and large-scale RSS variations are extracted to indicate on- or off-body radio propagations. Compared with the prior work, our system presents two main differences. First, along with time domain radio features, frequency domain features are abstracted to give a comprehensive description of on- and off-body radio propagations. Second, our work develops a customized adversarial network to essentially extract underlying propagation patterns and obtains a better generalization performance under various user motions in diverse environments. 

\textbf{Wireless sensing with machine learning.} Despite other applications with radio frequency (RF) signals \cite{xiao2019sensor,he2019state}, machine learning approaches have been widely applied in wireless sensing tasks. In \cite{clancy2011}, unsupervised learning approaches are used to facilitate signal classification in spectrum sensing. A support vector machine is exploited to classify human motions based on RF characteristics \cite{geng2016enlighten}. In \cite{shi2017smart}, a deep learning based user authentication scheme is proposed by using Wi-Fi signals, which capture unique human physiological and behavioral characteristics that are inherited from daily activities. Furthermore, relying on wireless signals, 2D and 3D human poses are estimated through walls and occlusions with the usage of cross-modal networks in \cite{zhao2018through,zhao2018rf}, respectively. In this work, we extract PHY signatures existing in radio signals and input them into a neural network to determine whether the signals are transmitted from on-body wireless devices.

\textbf{Adversarial learning.} Our system adopts an adversarial neural network for wireless device authentication. The adversarial network is originally proposed to estimate the density of an unknown distribution of digital images in \cite{goodfellow2014generative}. Thereafter, it is applied to promote the generalization performance of deep neural networks for predictive tasks. A semi-supervised model \cite{ganin2016domain} is trained through a domain adversarial training for image classification. Moreover, an adversarial multi-task model \cite{shinohara2016adversarial} is developed for robust speech recognition. In \cite{zhao2017learning}, a conditional adversarial model is introduced for sleep stage prediction. In this work, a customized adversarial network is developed to eliminate irrelevant information on user motions and surrounding environments, and ultimately to boost the performance of on-body IoT device authentication.

\section{Exploiting PHY Signatures in BAN Channels} \label{sec: motivation}

\subsection{Threat Model}

On-body devices, which are colocated with the wearable device on the same body, are considered to be legitimate. In contrast, attackers are off-body devices, which are not carried by the same user. They may locate at another user or somewhere else to actively broadcast malicious messages. We do not consider attacks from on-body devices, because it is normal for a user to check the ownership of an IoT device before wearing it. Moreover, we do not take into account passive attacks, i.e., eavesdropping attacks. In addition, attackers may be equipped with advanced hardware and have been aware of the transmission technology and deployed security mechanism. In this situation, they can forge the MAC addresses of valid devices and inject fake data into the network.

\subsection{Theoretical Explanation of Distinct On- and Off-Body Radio Propagations}
Since the human body is basically a low-loss dielectric at microwaves frequencies, including Wi-Fi and Bluetooth frequency bands, radio propagations between two on-body devices are significantly influenced by the user's body. Previous measurements \cite{ryckaert2004channel,di2011body} have demonstrated that creeping waves, which are diffracted by human tissues and spread out along the human body, play a predominant role in on-body electromagnetic wave propagations. According to creeping wave theory \cite{alves2011analytical}, the electric field over the conducting surface for the vertical polarization on the elliptical path can be expressed as
\begin{align} \label{model of creeping waves}
\mathbf{E} = 2 \sqrt{\frac{\eta}{2 \pi}} \frac{\sqrt{P_{Tx}G_{Tx}}}{d} e^{-jkd}L\left( a, b, \phi, \varphi \right),
\end{align}
where $ d $ is the distance between Tx and Rx antennas, $ \eta $ the vacuum wave impedance, $ P_{Tx} $ the transmission power, $ G_{Tx} $ the gain of the Tx antenna, and $ k $ the wave number in the free space. Moreover, $ L(\cdot) $ represents the attenuation factor that indicates the loss on the surface, and it is a function of $ a $ and $ b $, i.e., the semi-major and semi-minor axises of the ellipse respectively, $ \phi $ the exit point angle at Tx, and $ \varphi $ the trapping point angle at Rx. Furthermore, compared with the vertically polarized component, the horizontal component suffers more attenuation. Thus, the orientations of on-body antennas also have a great impact on the path loss of creeping waves.

Eq.~\ref{model of creeping waves} suggests that the body surface and the positions and orientations of Tx and Rx antennas, rather than environmental dynamics, dominate the attenuation of on-body propagations. Specifically, when two transceivers are both deployed on the same human body, any body movement can change the body surface as well as antenna positions, which consequently cause variations in the Tx-Rx distance $ d $ and the attenuation factor $ L(\cdot) $. As a result, on-body signals would be stable in a static motion status, and they will fluctuate dramatically when the body moves.

On the contrary, radio waves between a pair of devices that are not placed on the same body typically propagate in a different manner. Off-body signals are usually reflected by surrounding floor, walls and furniture (small-scale fading) and disturbed by Tx-Rx distance changes \cite{tse2005fundamentals} (large-scale fading). Compared with on-body signals, off-body signals are mainly comprised of LOS and multi-path components, and are less sensitive to the changes of the body surface and antenna positions. Therefore, we see that distinct propagation patterns exist between on- and off-body radio waves.

\subsection{Feasibility Study}

\begin{figure}
	\centering
    \subfigure[Standing in an indoor environment.]{
	\includegraphics[width=0.31\linewidth]{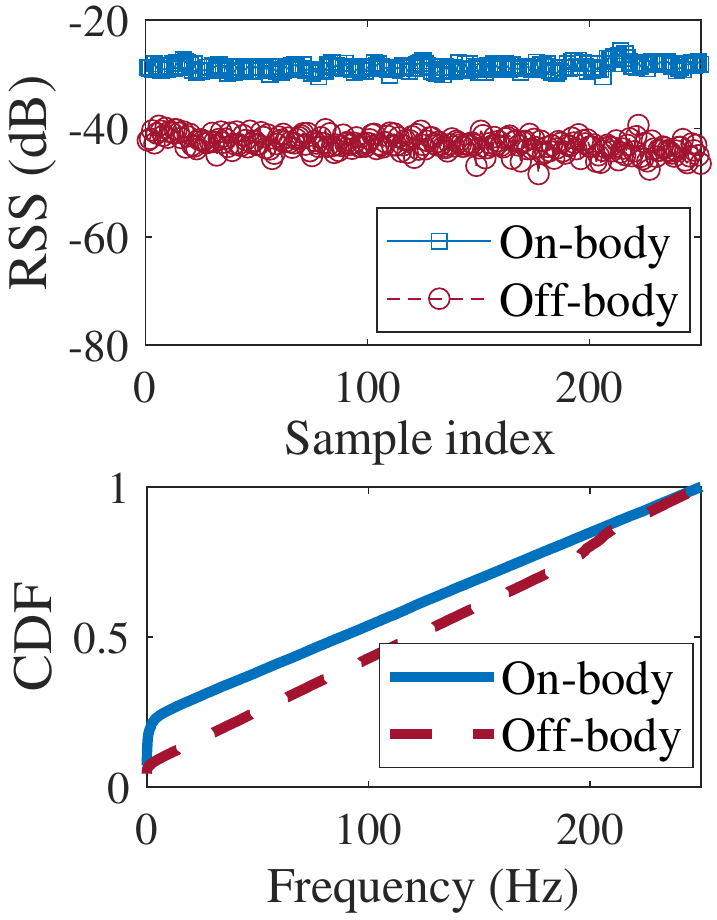}}
    \subfigure[Walking in an indoor environment.]{
	\includegraphics[width=0.31\linewidth]{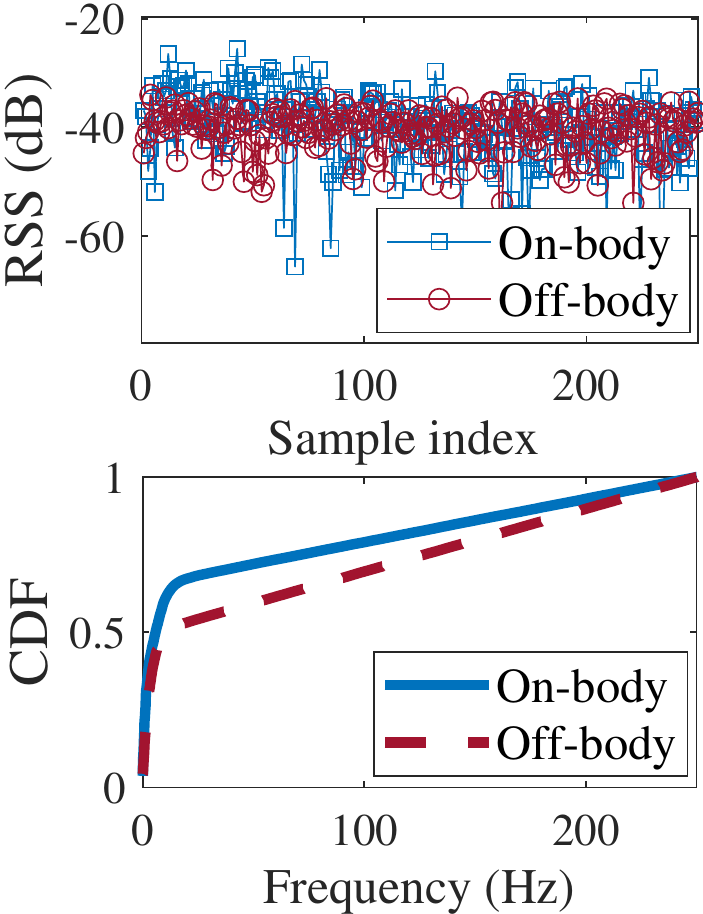}}
    \subfigure[Standing in an outdoor environment.]{
	\includegraphics[width=0.31\linewidth]{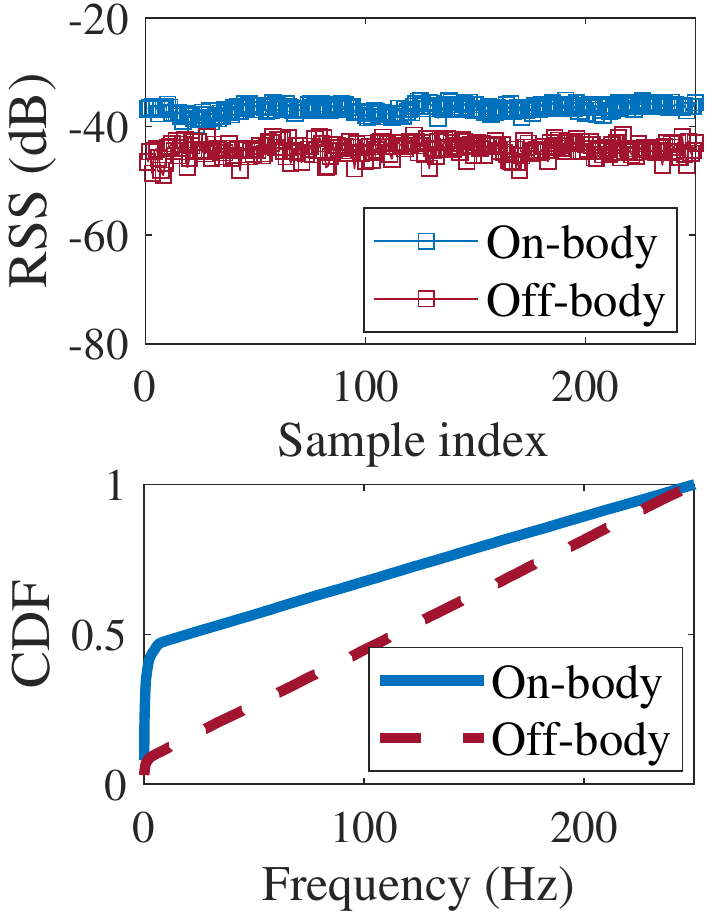}}
    \caption{RSS and CDF of on- and off-body radio signals in different scenarios.}
	\label{comparisons between time and frequency domain} 
\end{figure}

Based on the above analysis, we conduct a motivational experiment to demonstrate the feasibility of exploiting radio propagation features to verify on-body IoT devices. In the experiment, two USRP devices are carried by a volunteer to work as a pair of on-body Tx and Rx. The left device is placed on another volunteer, acting as an off-body Tx. We collect on- and off-body signals in three different scenarios, i.e., standing and walking in an indoor environment, respectively, and standing in an outdoor environment. Fig.~\ref{comparisons between time and frequency domain} depicts the RSS and cumulative distribution function (CDF) of the collected signals. We observe that compared with off-body signals, on-body signals are more stable when the user stands still, while having a lager RSS variance in the walking status. This observation testifies that on-body propagations are highly sensitive to user body motions. Moreover, off-body signals always fall into the high frequency range with a higher probability in comparison with on-body signals in each scenario, which verifies that off-body propagations are more susceptible to environmental dynamics. 

The above experimental observations verify that differentiable radio propagation patterns exist between on- and off-body channels in each scenario. This supports our premise that we can rely upon PHY signatures to authenticate various on-body IoT devices. 

\section{Adversarial Network Based Device Authentication}\label{sec: overview}

%In this section, we first discuss the design rationale of using adversarial networks. Next, we present an overview of our authentication system. Finally, we describe how to integrate our PHY-based device authentication system with upper-layer security protocols.

\subsection{Design Rationale}
It is, however, non-trivial to reliably capture propagation patterns from real-world radio traces. As shown in Fig.~\ref{comparisons between time and frequency domain}, although on- and off-body signals show distinguishable propagation patterns in each scenario, their patterns are remarkably different between the three scenarios. Consequently, an authentication model that is trained under a specific user motion in a specific environment will typically not generalize well in different scenarios. 

To deal with such dilemma, we resort to adversarial neural networks, which have recently surfaced as a popular tool to discover transferable features in the deep learning field and have proven their advantages in many real-world applications \cite{ganin2016domain,zhao2017learning,shinohara2016adversarial}. Being a branch of deep learning approaches, adversarial networks facilitate automatic extraction of complex and latent feature representations by adopting a hierarchical structure \cite{goodfellow2016deep}. More importantly, different from traditional approaches that learn transferable features, such as autoencoders \cite{kramer1991nonlinear}, adversarial networks have the ability to eliminate irrelevant features in learned representations with an adversarial training criterion. Specifically, in the application of on-body authentication, user body motions and surrounding environments can easily incur different levels of variances and dynamics in RSS measurements. Once these noisy measurements are fed into a model for training, motion and environment specific features will be learned, which consequently hampers its authentication performance  at the testing phase. Therefore, we reap the benefits of adversarial networks to exclude irrelevant features induced by motions and environments and further recognize underlying on- and off-body propagation patterns in real-world scenarios. Towards this end, we propose an adversarial network based authentication system to seamlessly authenticate various on-body IoT devices.

\subsection{Design Overview}

\begin{figure}
	\centering
	\includegraphics[width=0.95\linewidth]{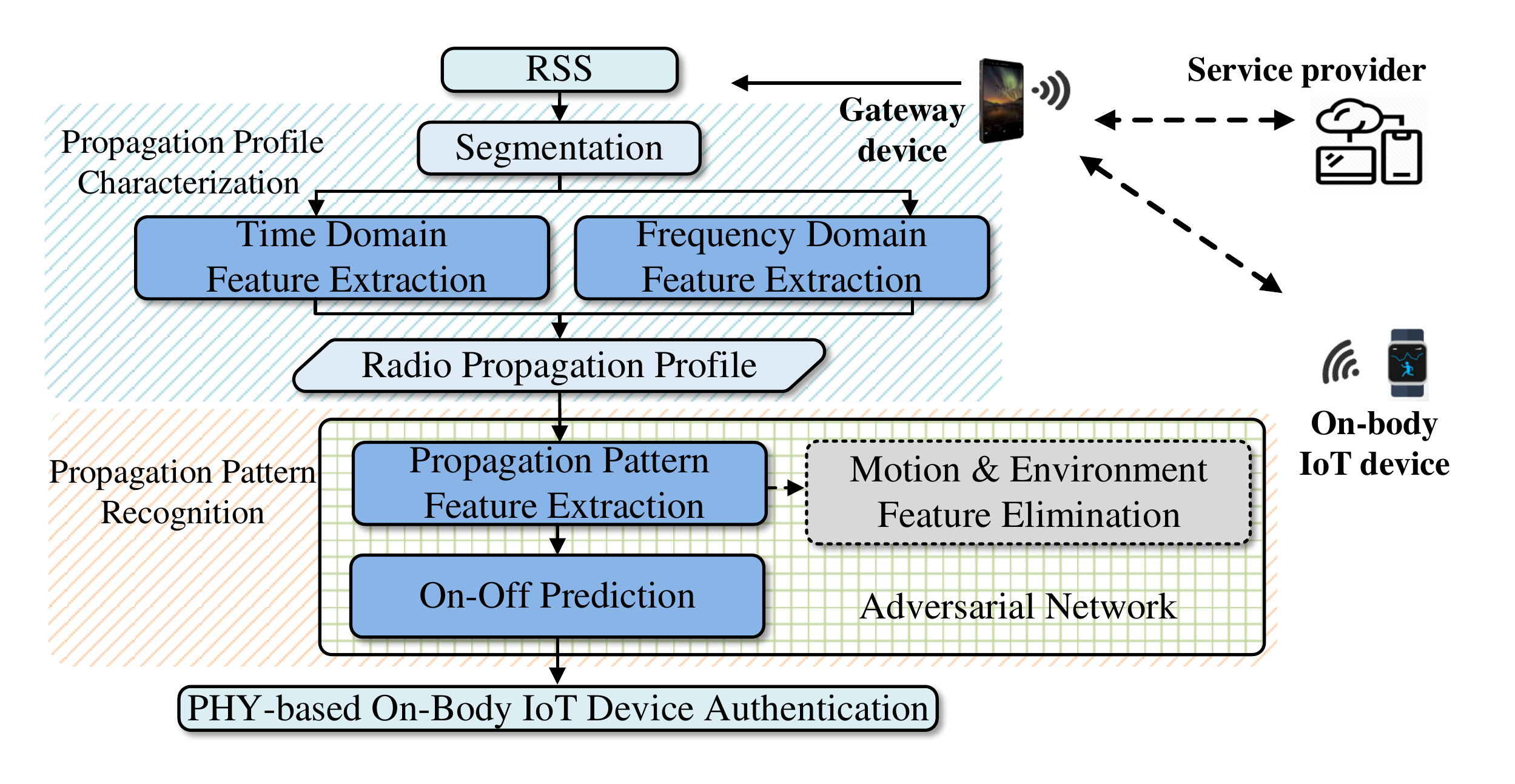}
	\caption{System overview. The dashed gray component exists only in the training phase.}
	\label{fig:system-flow}
\end{figure}

Our system takes advantage of an adversarial network to extract distinct radio propagation patterns for on-body device authentication. Fig.~\ref{fig:system-flow} illustrates the framework of our authentication system. It takes as input RSS time series and outputs the corresponding device authentication results. It is worth noting that to verify RSS measurements of various low-end embedded IoT devices, our authentication system runs on gateway devices, such as smartphones, which have sufficient capability to perform low-latency and accurate learning based inferences \cite{mobilenet}. 

The core of our authentication system includes two components -- \textit{Propagation Profile Characterization} and \textit{Propagation Pattern Recognition}.
\begin{enumerate}
    \item  \textbf{Propagation profile characterization.} First, this component divides the RSS time series into multiple basic segments. Then, representative time and frequency domain features are extracted for fine-grained characterization of potential propagation patterns. Finally, the extracted features are integrated into radio propagation profiles for future pattern recognition by the adversarial network.
	\item  \textbf{Propagation pattern recognition.} Upon receiving a propagation profile, the adversarial network first utilizes a functional block to abstract a feature representation in terms of on- and off-body propagations. Subsequently, the network infers the identity of a connected IoT device through an on-off prediction block. Moreover, an adversarial block is added to eliminate motion and environment specific features in the feature representation in the training phase. All blocks are learned through an adversarial training process to promote the emergence of features that are resilient to motion variances and environment changes. 
\end{enumerate}

\subsection{Integration with Upper-Layer Security Protocols}

Based on PHY signatures, our authentication system can integrate with existing security protocols in the upper layers to shield human-centric IoT networks from active attackers. Integrating our system with the 802.11 protocol, the final cross-layer protocol not only follows similar reasoning with upper-layer security standards but also takes the propagation patterns in PHY into consideration. Specifically, we shed light on how our system can be exploited to secure IoT device pairing and data transmission against authenticated spoofing attacks and authentication deadlock attacks, respectively.

\begin{figure}
	\centering
	\includegraphics[width=0.85\linewidth]{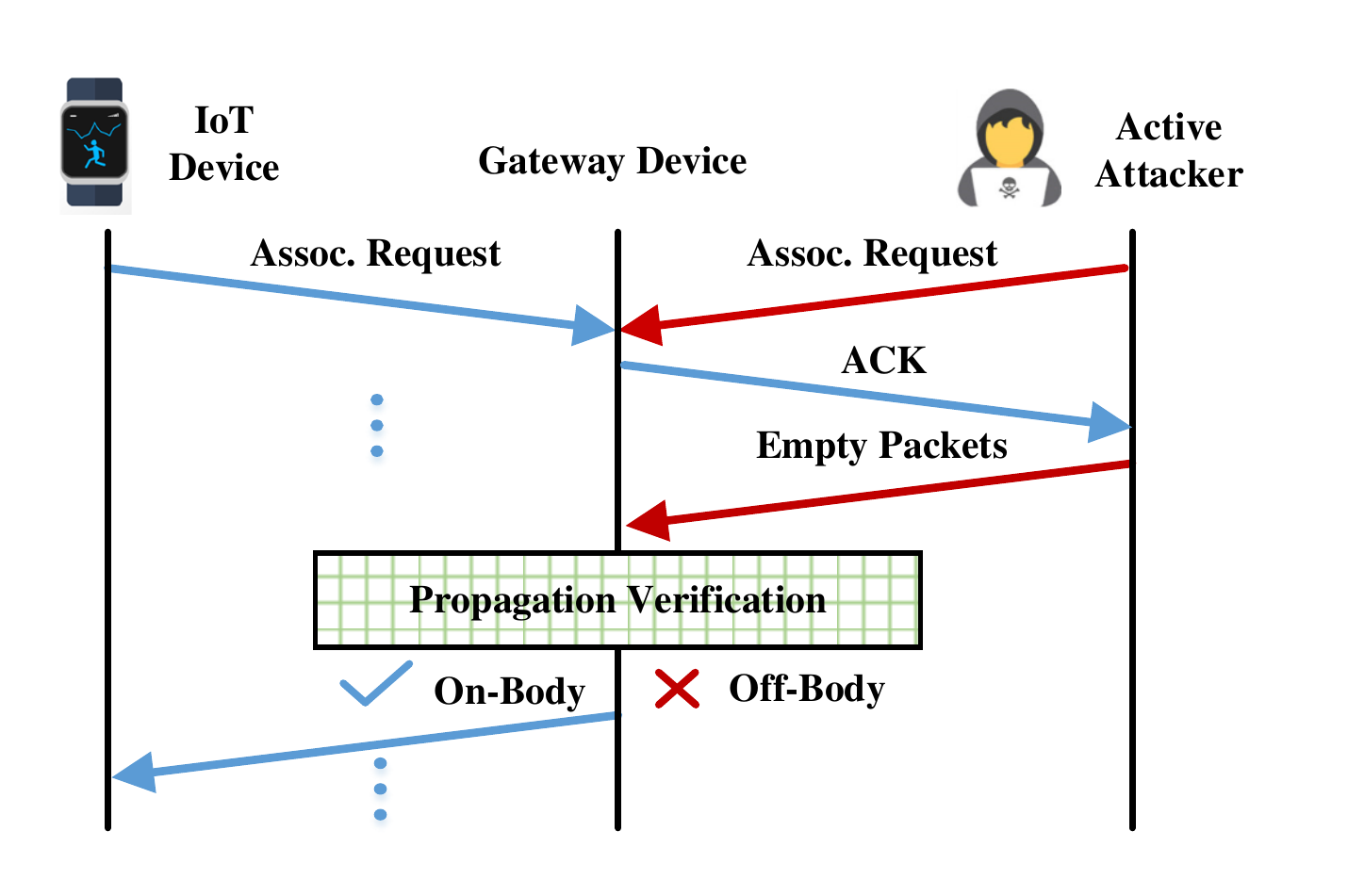}
	\caption{Authenticated spoofing attack and mitigation.}
	\label{fig:spoofing attack}
\end{figure}

\begin{figure}
	\centering
	\includegraphics[width=0.85\linewidth]{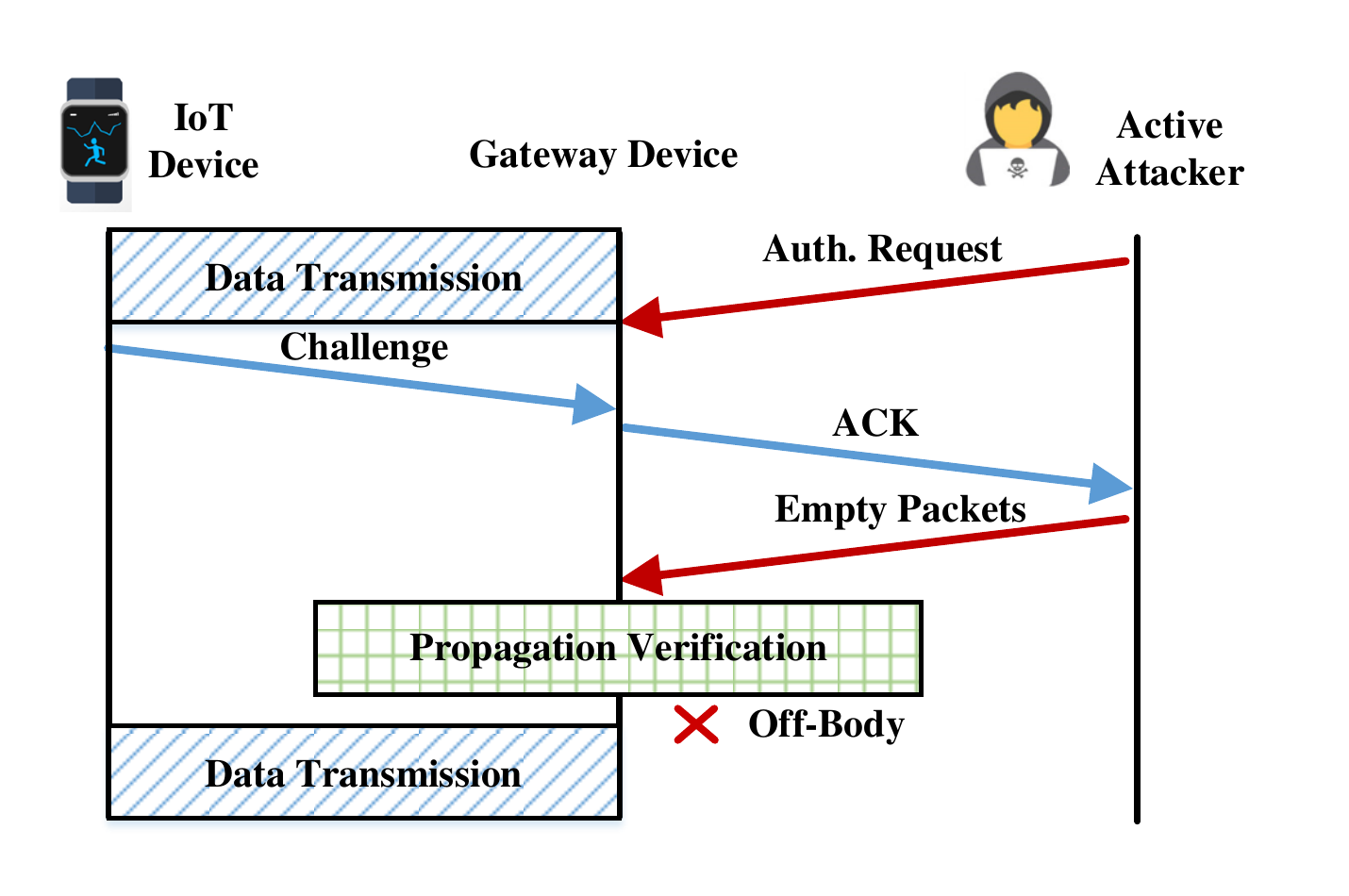}
	\caption{Authentication deadlock attack and mitigation.}
	\label{fig:deadlock attack}
\end{figure}

\textbf{Authenticated spoofing mitigation.} In many cases, users' authenticated login credentials and MAC addresses are susceptible to malicious attackers. Once deciphering this confidential information, an attacker can associate with a gateway device by masquerading a legitimate device and thereafter launch a variety of spoofing attacks on the IoT system. For instance, it can either inject fake messages into the system or steal users' personalized profiles from it.

Our system can mitigate these types of attacks by consolidating upper-layer security protocols with PHY propagation verification in the device association process as described in Fig.~\ref{fig:spoofing attack}. Specifically, when hearing an association request message from a surrounding device, the gateway device sends back an acknowledgment frame (ACK) to request propagation pattern verification. In response to the ACK, the surrounding device must transmit a series of empty packets to the gateway device. Subsequently, the gateway device decides whether the transmitter is carried by the same user based on our authentication system. If the propagation pattern is recognized to be on-body, the gateway device regards it as an authorized IoT device and starts subsequent communication links. Otherwise, the gateway device deems it as a malicious attacker and denies the association request.

\textbf{Authentication deadlock mitigation.} Authentication deadlock attack is one of Denial-of-Service (DoS) attacks. In 802.11 protocols, a legitimate IoT device must be authenticated and associated with a gateway device before data transmission process. During data transmission between the IoT device and gateway device, an attacker injects an Open System Authentication Request frame to the gateway device in the name of the IoT device. This attack will consequently lead to an authentication deadlock, causing the gateway device to delete the authenticated association with the IoT device and thus cannot transmit or receive any frames for the victim device for a few minutes \cite{eian2011modeling}.

Our security protocol defends against authentication deadlock attacks with slight changes in existing upper-layer protocols as shown in Fig.~\ref{fig:deadlock attack}.Upon receiving an authentication request during data transmission, the gateway device detects that the received request nominally comes from an device that is actually in the authenticated state, and thus it can consider this authentication frame as suspicious \cite{Xiong2013secure,eian2012formal,singh2015ieee}. Then, the gateway device will not delete the authenticated association immediately but allow the IoT device to send a challenge frame denying that the request is from itself. After that, the gateway device sends an ACK frame to the requesting node to start a propagation verification for incoming empty signals. If the propagation pattern mismatches the on-body one, the gateway device decides to drop the authentication request and continues with the previous data transmission.

\section{Propagation Profile Characterization}\label{sec: profile}

\begin{figure}
	\centering
	\includegraphics[width=0.95\linewidth]{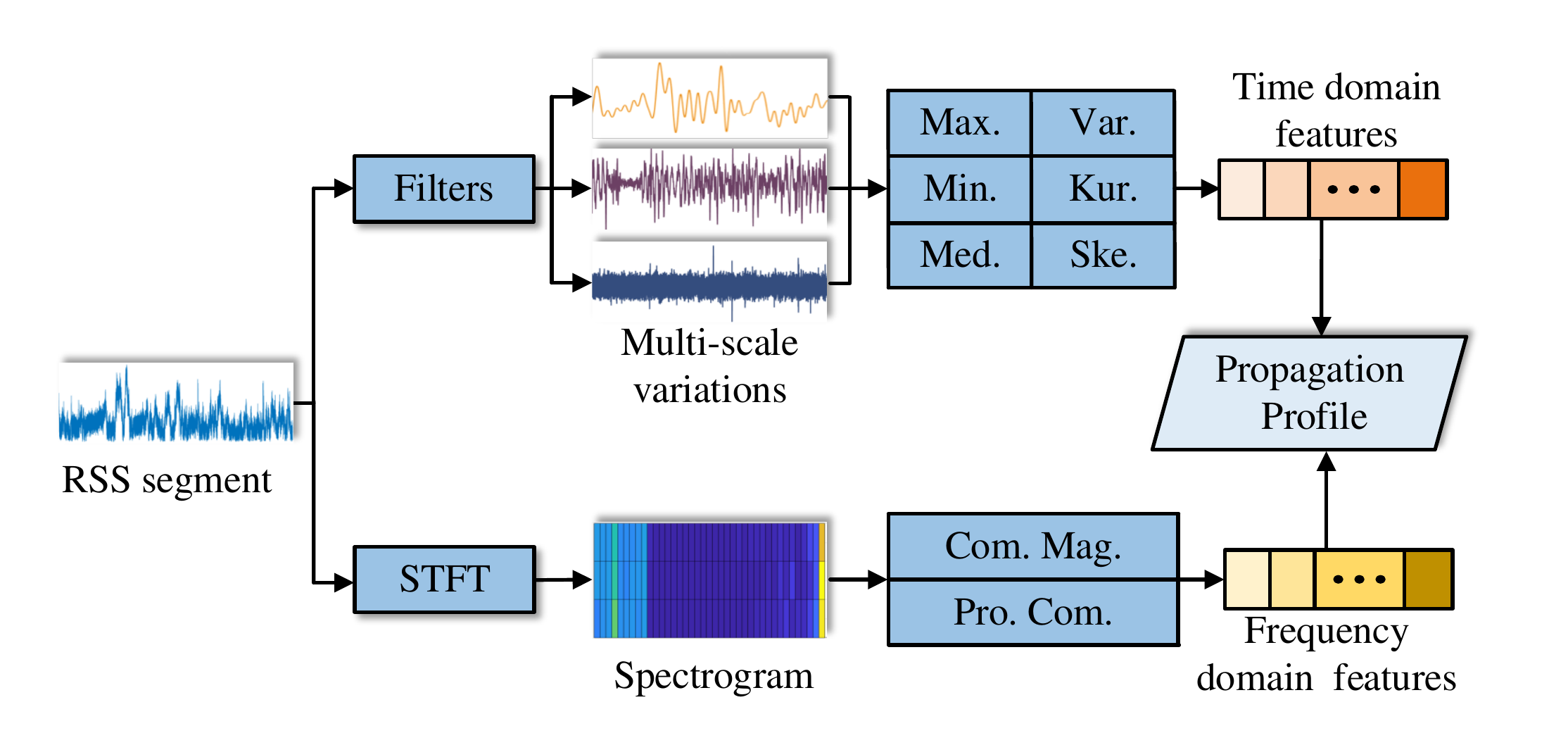}
	\caption{Work flow of the propagation profile characterization.}
	\label{fig:propagationprofilecharacterization}
\end{figure}

%In this section, we describe how to construct radio propagation profiles from noisy real-world radio traces. As shown in Fig.~\ref{fig:propagationprofilecharacterization}, it starts with segmenting RSS time series into basic units and then extracts reliable radio features from both the time and frequency domains of RSS segments.

\subsection{Signal Segmentation} 

Our system first partitions RSS measurements into multiple segments. As an RSS segment is a basic unit for device authentication, the segment interval needs to be carefully determined. If the interval is too long, on- and off-body signals will be probably both included in the same segment. If it is too short, the system will be unable to recognize any segment. We empirically find that a time interval of 5s is capable of correctly differentiating over $ 90\% $ of on- and off-body IoT devices.

\subsection{Time Domain Feature Extraction} 

Since on- and off-body signals have different levels of impact from body motions, large- and small-scale fading, we first decompose each RSS segment into multi-scale variations by using filters. As creeping waves are sensitive to body motions and their frequencies fall into relatively low frequency bands with a high probability \cite{xiong2006hand}, a band-pass filter is leveraged to extract motion-induced variations. Based on our experimental observations, most fluctuations caused by body motions fall between 0.5 Hz and 15 Hz. Variations in the residual low and high frequency bands are also extracted by a low-pass filter and a high-pass filter, respectively, as large- and small-scale variations.

\begin{figure}%[t] 
	\centering
	\subfigure[Maximum.]{
		\includegraphics[width=0.4\linewidth]{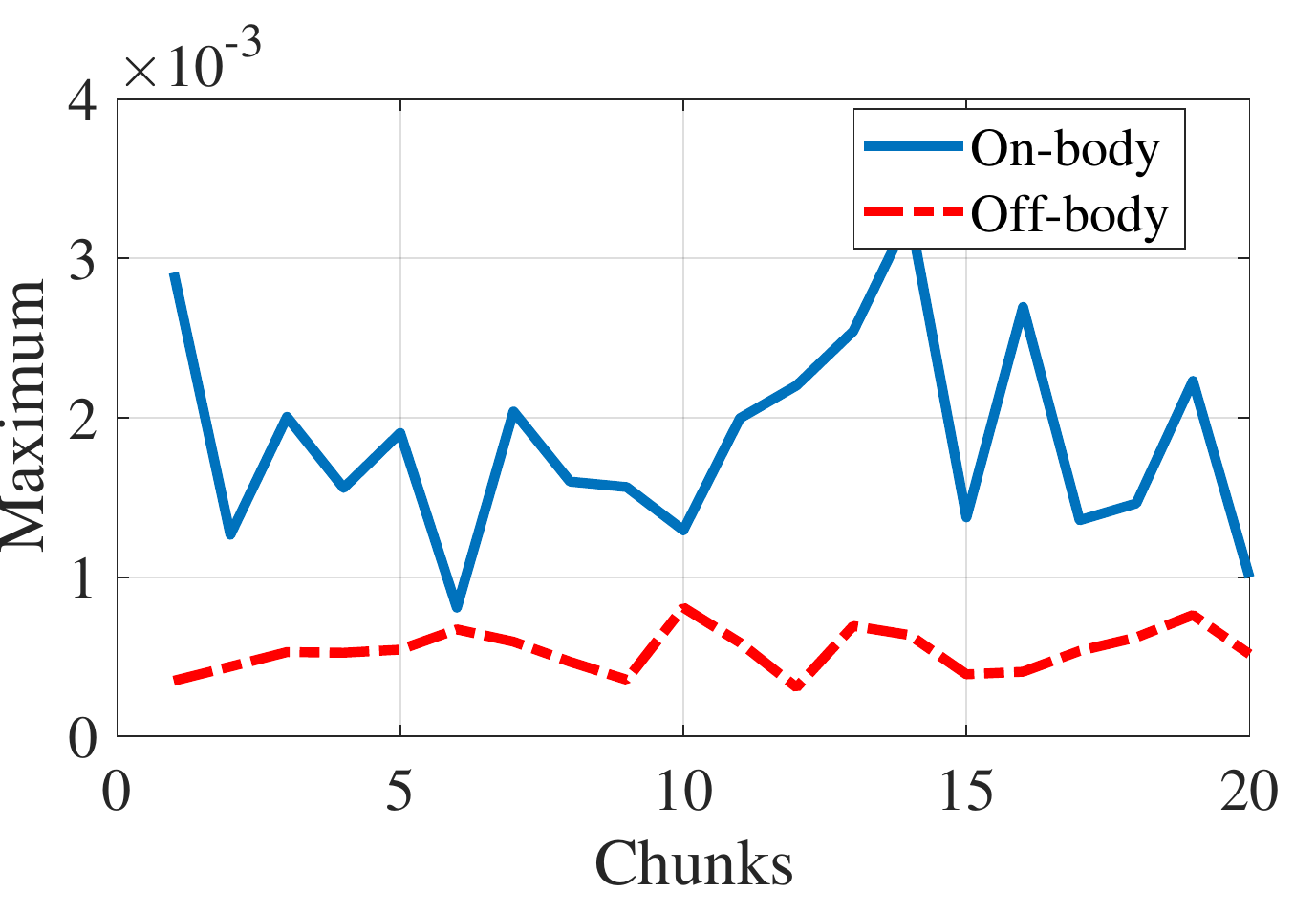}}
	\label{2a}
	\hspace{0.3cm}
	\subfigure[Minimum.]{
		\includegraphics[width=0.4\linewidth]{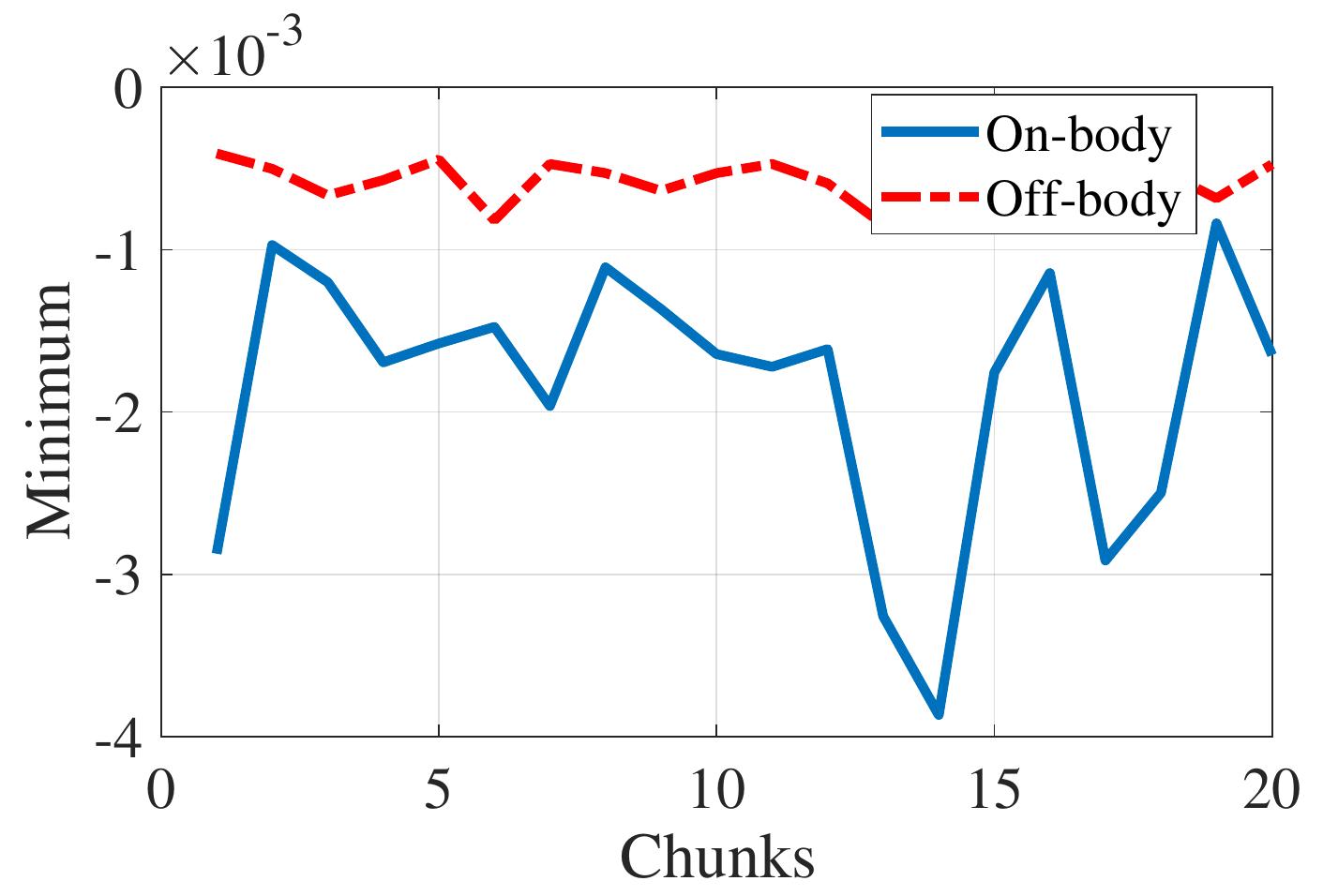}}
	\label{2b}\\
	\subfigure[Median.]{
		\includegraphics[width=0.4\linewidth]{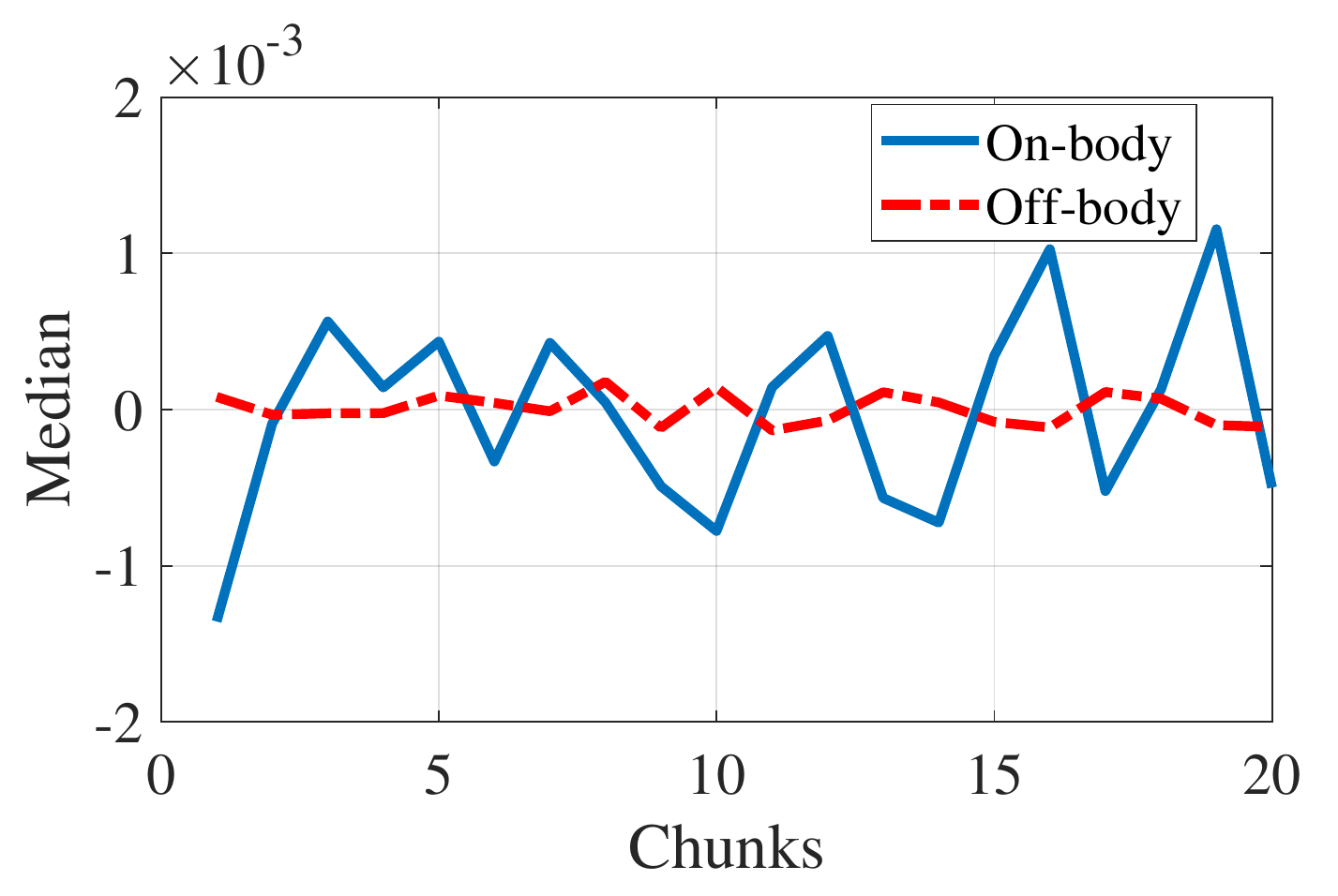}}
	\label{2c}
	\hspace{0.3cm}
	\subfigure[Variance.]{
		\includegraphics[width=0.4\linewidth]{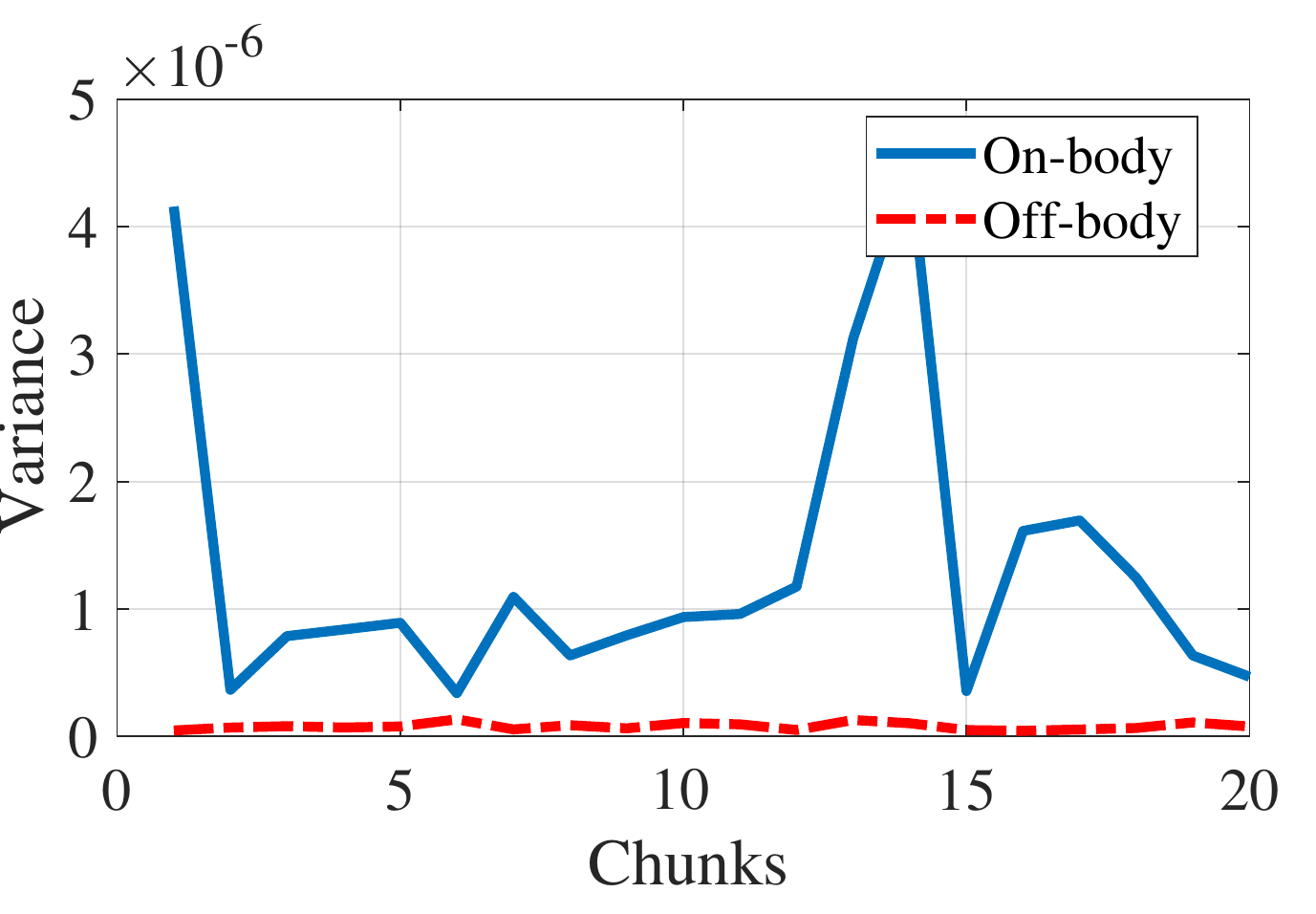}}
	\label{2d} 
	\caption{Examples of time domain features.}
	\label{time domain features} 
\end{figure}

With multi-scale variations, we select six time domain features, including \textit{maximum}, \textit{minimum}, \textit{median}, \textit{variance}, \textit{kurtosis} and \textit{skewness}, to characterize propagation signatures from each kind of variations. The maximum, minimum, median and variance are chosen to describe the impact from the human body, because dramatic body vibration typically contributes to rapid changes in the maximum, minimum and median and also results in a large variance. Kurtosis and skewness show the symmetry and asymmetry of radio signals, respectively, and can potentially capture propagation patterns due to the fact that both symmetric and asymmetric components are richly shared in radio waves. For finer-grained feature extraction, we divide each kind of variations into ten chunks and extract six features from each chunk. Therefore, a total of 180 feature points are extracted to describe radio propagation signatures from the time domain of an RSS segment. Fig.~\ref{time domain features} presents some time domain features extracted from motion-induced variations when an user stands still in a normal office setting. We can observe that these features are significantly different between on- and off-body signals, which encourages us to exploit these time domain features to characterize distinct BAN radio propagations.

\subsection{Frequency Domain Feature Extraction} 

Different power distributions on the frequency band between on- and off-body signals have been clearly presented in Fig.~\ref{comparisons between time and frequency domain}. Thus, besides time domain features, frequency domain features are also extracted to capture distinct signatures of different BAN radio propagations.

To abstract frequency domain features, we start by performing a Short-Time Fourier Transform (STFT) on each RSS segment to obtain its two-dimensional spectrogram. Specifically, with a signal sampling rate of 500 Hz, we conduct a 1000-point Fast Fourier Transform (FFT) within a 2s sliding window, shifting 1s each time to make full use of sampling data. To summarize information in the frequency domain, the frequency band of each spectrogram, i.e., [0,250] Hz, is partitioned into 40 intervals, each of which is associated with a frequency component of the segment. To effectively indicate propagation signatures, we equally segment the low frequency band, i.e., [0,15] Hz, into 30 intervals and the residual high frequency band into 10 intervals, and we sum up the magnitudes in each interval in every FFT result. In this way, we transform a two-dimensional spectrogram into a 4$ \times $40 matrix $ \mathbf{M} $. Then we take two frequency domain features from $ \mathbf{M} $: the \textit{component magnitude} (or each element in $ \mathbf{M} $) and the \textit{proportion of each component} (PC), such that $ PC(j)=\frac{\sum_{i=1}^{4} \mathbf{M}(i,j)}{\sum_{j=1}^{40} \sum_{i=1}^{4} \mathbf{M}(i,j)} $, where $ j=1, \cdots, 40$. Finally, a total of 200 feature points are extracted from the frequency domain of an RSS segment. 

\begin{figure}
	\centering
	\includegraphics[width=0.95\linewidth]{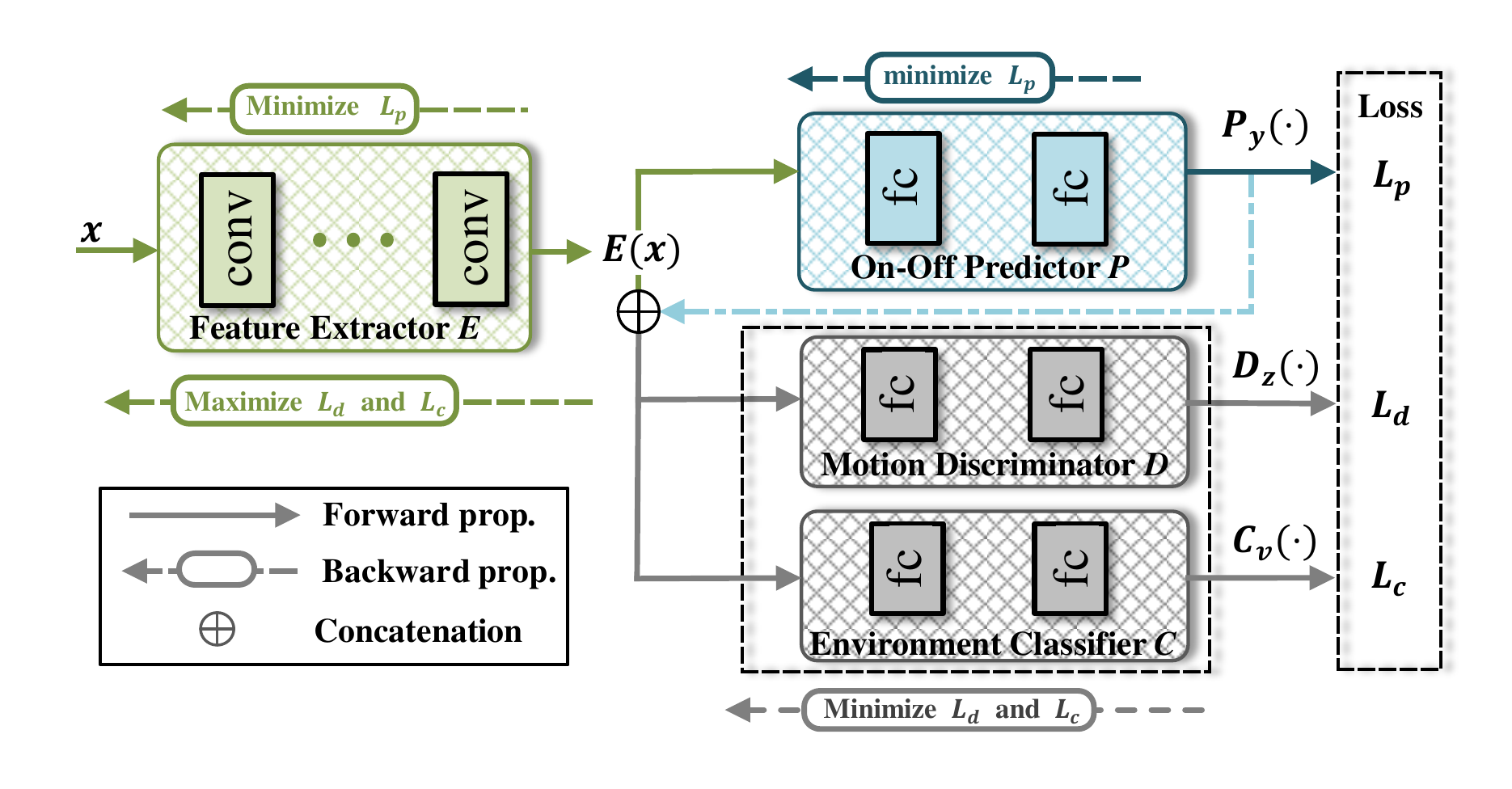}
	\caption{Our adversarial multi-player network and corresponding adversarial training criterion. The model encompasses a Feature Extractor $ E $, an On-Off Predictor $ P $, a Motion Discriminator $ D $, and an Environment Classifier $ C $.}
	\label{fig:adversarialmodel}
\end{figure}

\section{Propagation Pattern Recognition}\label{sec: recognition}

%In this section, we elaborate on the propagation pattern recognition. First, we customize an adversarial multi-player network to distinguish between on- and off-body IoT devices under different user motions in diverse environments. Then, we theoretically analyze the customized network and prove its effectiveness. Next, we design an adversarial training algorithm.

\subsection{Adversarial Model}

After the profile characterization, we consider the propagation pattern recognition as a binary classification task $ \left( \mathcal{X}, \mathcal{Y} \right) $, where $ \mathcal{X} \subseteq \mathbb{R}^{n} $ is the sample space and $ \mathcal{Y} = \left\lbrace 0,1 \right\rbrace  $ the target label set. Specifically, each $ \mathbf{x} \in \mathcal{X} $ is a radio propagation profile sample, and $ y \in \mathcal{Y} $ indicates the corresponding on- or off-body device. Moreover, for each $ \mathbf{x} $, $ z \in \left\lbrace 1, 2, \cdots, n_{z} \right\rbrace  $ and $ v \in \left\lbrace 1, 2, \cdots, n_{v} \right\rbrace  $ denote a pair of auxiliary labels that refer to the motion and environment, respectively, that $ \mathbf{x} $ is sampled from.

For effective classification, we develop an adversarial multi-player neural network, as shown in Fig.~\ref{fig:adversarialmodel}. In particular, our model consists of four components -- a \textit{Feature Extractor $ E $}, an \textit{On-Off Predictor $ P $}, a \textit{Motion Discriminator $ D $} and an \textit{Environment Classifier $ C $}. 

\textbf{Feature extractor $ E $.} We leverage a convolutional neural network (CNN) to aggregate information over time and frequency domain features to extract underlying radio propagation patterns. More specifically, eight 1D convolutional layers are stacked in our feature extractor. At each layer of $ E $, 128 convolutional kernels with the kernel size 1$\times$3, stride 1 and padding 0 are used to filter valuable ingredients from the previous layer's output. In addition, we use a max-pooling (MP) layer with the kernel size of 2 to reduce the representation size and a rectified linear unit (ReLU) to introduce nonlinearity into the model. Thus, at layer $ l $, a latent representation $ \mathbf{r}^{l} $ are computed as
\begin{align}
\mathbf{r}^{l} = \sigma_{MP + ReLU}  \left( \mathbf{f}^{l} * \mathbf{r}^{l-1} \right), 
\end{align}
where $ * $ is the convolution operator, $ \mathbf{f}^{l} $ convolutional kernels and $ \mathbf{r}^{l-1} $ the output of layer $ l-1 $. At last, given an input sample $ \mathbf{x} $, the corresponding feature representation can be obtained by
\begin{align}
E(\mathbf{x}) = CNN(\mathbf{x};\Theta_e),
\end{align}
where $ \Theta_e $ denotes the extractor's trainable parameters.

\textbf{On-off predictor $ P $.} Based on a feature representation $ E(\mathbf{x}) $, we parameterize the on-off predictor as a fully-connected neural network. In particular, two fully-connected layers with Sigmoid and Softmax functions, respectively, are used to map $ E(\mathbf{x}) $ into a two-dimensional probability vector $ P_{y} (\cdot|E(\mathbf{x})) \in \mathbb{R}^{2} $ in terms of on- and off-body devices, as follows
\begin{align}
\mathbf{g} = \sigma_{Sigmoid} (\mathbf{w}_g \cdot E(\mathbf{x})+\mathbf{b}_g),
\end{align} 
\begin{align}
P_{y} (\cdot|E(\mathbf{x})) = \sigma_{Softmax} (\mathbf{w}_h\cdot \mathbf{g}+\mathbf{b}_h),
\end{align} 
where $ \Theta_p =(\mathbf{w}_g,\mathbf{b}_g,\mathbf{w}_h,\mathbf{b}_h) $ is the predictor's parameters. Once the probability vector $ P_{y} (\cdot|E(\mathbf{x})) $ is obtained, we can have a predicted target label $ \hat{y} $ for $ \mathbf{x} $, as follows
\begin{align}
\hat{y} = \arg \mathop{\max} \limits_{y} P_{y} (\cdot|E(\mathbf{x})).
\end{align}

\textbf{Adversarial discriminator $ D $ and classifier $ C $.} Since ongoing body motions and surrounding environments have different impacts on radio signals, two adversaries $ D $ and $ C $ are adopted to remove their respective features in the feature representation $ E(\mathbf{x}) $. Specifically, body motions typically cause Tx-Rx distance changes or shadowing, resulting in large-scale variations of radio signals, however surrounding environments incur rich multipath propagations and small-scale variations. Thus, the information corresponding to motions and environments in $ E(\mathbf{x}) $ can be considered to be independent. Note that the adversarial components will not increase the computational complexity when our system performs on-body authentication, because they are only needed in the training phase.

Since simply wiping out all dependencies between the feature extractor and adversaries could degrade the performance of target label prediction, our model adopts a conditional adversarial architecture \cite{zhao2017learning} for better generalization performance. For this purpose, we concatenate the outputs of feature extractor and on-off predictor as the input of two adversaries. Hence, $ D $ and $ C $ predict body motions and environments with outputs $ D_{z} \left(\cdot \big| E(\mathbf{x}), P_{y} (\cdot|E(\mathbf{x})) \right) \in \mathbb{R}^{n_z}  $ and $ C_{v} \left(\cdot \big| E(\mathbf{x}), P_{y} (\cdot|E(\mathbf{x})) \right) \in \mathbb{R}^{n_v} $. In the proposed model, we also parameterize the discriminator and classifier as two fully-connected neural networks, which both have the same configurations with the predictor. Moreover, the parameters of $ D $ and $ C $ are denoted as $ \Theta_d $ and $ \Theta_c $, respectively.

\textbf{Adversarial training criterion.} To obtain useful network parameters, we train our adversarial model on a set of training data, which obeys the distribution $ q_{train} (\mathbf{x}) $. For the predictor $ P $, the cross-entropy is used to calculate the discrepancy between the prediction $ P_{y} \left( \cdot|E(\mathbf{x})\right)  $ and the true posterior distribution $ q_y (\cdot|\mathbf{x}) $ over $ q_{train} (\mathbf{x}) $, as follows
\begin{align}
\mathcal{L}_P (P,E) \triangleq \mathbb{E}_{\mathbf{x} \sim q_{train} (\mathbf{x})} \mathbb{E}_{y \sim q (y|\mathbf{x})} \left[ -\log P \left( y | E(\mathbf{x})\right) \right].
\end{align}
By minimizing $ \mathcal{L}_P $, the parameters $ \Theta_e $ and $ \Theta_p $ can be updated.

Moreover, we define the loss of $ D $ as the cross-entropy between its output $ D_z \left(\cdot \big| E(\mathbf{x}), P_y (\cdot|E(\mathbf{x})) \right) $ and the true conditional distribution $ q_z (\cdot|\mathbf{x}) $ over $ q_{train} (\mathbf{x}) $, which is expressed as
\begin{align}\label{loss of d1}
\mathcal{L}_{D} \left( D,E;P\right) \triangleq \mathbb{E}_{\mathbf{x} \sim q_{train} (\mathbf{x})} \mathbb{E}_{z \sim q (z|\mathbf{x})} \left[ -\log D \left( z \big| E(\mathbf{x}),P_y (\cdot|E(\mathbf{x})) \right) \right].
\end{align}
Similarly, the loss of $ C $ is given by
\begin{align}
\mathcal{L}_{C} \left( C,E;P\right) \triangleq \mathbb{E}_{\mathbf{x} \sim q_{train} (\mathbf{x})} \mathbb{E}_{v \sim q (v|\mathbf{x})} \left[ -\log C \left( v \big| E(\mathbf{x}),P_y (\cdot|E(\mathbf{x})) \right) \right].
\end{align}
Note that to effectively train our multi-player model, the concatenation branch of the predictor's output is a one-way link (i.e., the dashed blue arrow line in Fig.~\ref{fig:adversarialmodel}), along which gradients of the adversarial components don't propagate back. Hence, the parameters $ \Theta_e $, $ \Theta_d $ and $ \Theta_c $ can be refined through the optimization of $ \mathcal{L}_{D} $ and $ \mathcal{L}_{C} $. 

Now that we have defined all loss functions, we proceed to implement an adversarial training criterion on our multi-player model for robust device authentication under various body motions in different environments. The key idea is that to generalize well in unseen scenarios, a predictive model is able to discriminate well between on- and off-body devices, but it cannot distinguish scenarios associated with input samples. To achieve this goal, we use a minimax game between $ E $, $ P $, $ D $ and $ C $ in the training phase. Particularly, $ P $, $ D $ and $ C $ aim to minimize their own losses for good prediction performance. However, $ E $ tries its best to maximize $ \mathcal{L}_{D} $ and $ \mathcal{L}_{C} $ to cheat its adversaries $ D $ and $ C $, respectively, and at the same time, it cooperates with $ P $ to minimize $ \mathcal{L}_{P} $. Through this minimax game, the multi-player model can finally learn transferable features that are resilient to body motions and environments.

According to the above objectives, we integrate all loss functions into one value function, which is given as  
\begin{align}\label{value function}
	 \mathcal{V}\left( E,P,D,C\right)  \triangleq & \mathcal{L}_P- \alpha \mathcal{L}_{D} - \beta \mathcal{L}_{C},
\end{align}
where $ \alpha > 0 $ and $ \beta > 0 $ are hyperparameters. Based on the value function~\eqref{value function}, the adversarial training criterion can be implemented by optimizing the following minimax problem:
\begin{align}\label{minimax training criterion}
\mathop{\min} \limits_{E,P} \mathop{\max} \limits_{D,C} \mathcal{V}\left( E,P,D,C\right).
\end{align}

\subsection{Theoretical Analysis of Adversarial Model}
In this subsection, we prove that the output of the on-off predictor becomes invariant to motion variances and environmental dynamics through the minimax game. Specifically, we first present the optimal predictor and adversaries in Proposition 1 and Proposition 2, respectively, without proving them, and refer the reader to \cite{zhao2017learning} (Proposition 2) for details. Then, we illustrate the virtual training criterion, optimal extractor and optimal outputs, respectively, in Corollary 1, Proposition 3 and Corollary 2. Differing from the theoretical efforts in the prior work \cite{zhao2017learning}, our analysis focuses on a practical adversarial model.
\begin{proposition} 
	(Optimal predictor) For a fixed extractor $ E $, the output of the optimal predictor $ P^{*} $ over $ q_{train} (\mathbf{x}) $ achieves
	\begin{align} \label{optimal predictor}
	P^{*}\left( y|E(\mathbf{x})\right) =q(y|E(\mathbf{x})),
	\end{align}	
	and the loss of $ P^{*} $ is
	\begin{align} \label{loss of optimal predictor}
	\mathcal{L}_{P^{*}} (E) \triangleq \mathop{\min} \limits_{P} \mathcal{L}_P (P,E) = H(y|E(\mathbf{x})),
	\end{align}
	where $ H(\cdot| \cdot) $ denotes the conditional entropy function.
\end{proposition}

Note that given $ E $, the equality~\eqref{optimal predictor} indicates the maximal predictive capability that a predictor $ P $ can learn from the feature representation $ E(\mathbf{x}) $ over $ q_{train} (\mathbf{x}) $.
\begin{proposition} 
	(Optimal discriminator and classifier) Given any extractor $ E $ and any predictor $ P $, the optimal discriminator $ D^{*} $ and classifier $ C^{*} $ have their losses, respectively, as
	\begin{align} \label{loss of optimal D}
	\mathcal{L}_{D^{*}}(E;P)  \triangleq \mathop{\min} \limits_{D} \mathcal{L}_{D}(D,E;P)
	= H\left( z \big|E(\mathbf{x}), P_y (\cdot|E(\mathbf{x})) \right),
	\end{align}
    \begin{align} \label{loss of optimal C}
	\mathcal{L}_{C^{*}}(E;P)  \triangleq \mathop{\min} \limits_{C} \mathcal{L}_{C}(C,E;P)
	= H\left( v \big|E(\mathbf{x}), P_y (\cdot|E(\mathbf{x})) \right).
	\end{align}
\end{proposition}

With the optimal predictor, discriminator and classifier, we proceed to simplify the minimax training criterion~\eqref{minimax training criterion}.

\begin{corollary} (Virtual training criterion) If $ P $, $ D $ and $ C $ have enough capacity and are trained to be optimal over $ q_{train} (\mathbf{x}) $, the minimax optimization~\eqref{minimax training criterion} is equivalent to the minimization of a virtual value function $ \mathcal{V}(E) $, which is expressed as  
	\begin{align} \label{actual minimax training criterion}
	\notag \mathcal{V}(E) \triangleq & H \left( y |E(\mathbf{x}) \right) - \alpha H\left( z \big| E(\mathbf{x}),q_y(\cdot|E(\mathbf{x}) )  \right) \\ & -  \beta H\left( v \big| E(\mathbf{x}),q_y(\cdot|E(\mathbf{x}) )  \right). 
	\end{align}
\end{corollary}

\begin{IEEEproof}
	Considering the optimal predictor $ P^{*} $ in Proposition 1, we can rewrite the losses of the optimal discriminator $ D^{*} $ and optimal classifier $ C^{*} $ in Proposition 2, by substituting \eqref{optimal predictor} into \eqref{loss of optimal D} and \eqref{loss of optimal C}, respectively, as
	\begin{align} \label{new loss of optimal D}
	\mathcal{L}_{D^{*}}(E) = H\left( z \big|E(\mathbf{x}),q_y (\cdot|E(\mathbf{x})) \right),
	\end{align}
	\begin{align} \label{new loss of optimal C}
	\mathcal{L}_{C^{*}}(E) = H\left( v \big|E(\mathbf{x}),q_y (\cdot|E(\mathbf{x})) \right).
	\end{align}
	According to the equalities \eqref{loss of optimal predictor}, \eqref{new loss of optimal D} and \eqref{new loss of optimal C}, the initial value function \eqref{value function} can be simplified as the virtual version \eqref{actual minimax training criterion}. Thus, optimizing the minimax optimization \eqref{minimax training criterion} equals to minimizing $ \mathcal{V}(E) $. 
\end{IEEEproof}

Based on the virtual training criterion, we can obtain the optimal extractor by minimizing $ \mathcal{V}(E) $. 

\begin{proposition} (Optimal extractor) \label{optimal extractor} 
	If $ E $, $ P $, $ D $ and $ C $ have enough capability and are trained to be optimal over $ q_{train} (\mathbf{x}) $, any optimal extractor $ E^{*} $ satisfies 
	\begin{align}\label{property_one}
	H \left( y| E^{*} (\mathbf{x}) \right) = H \left( y|\mathbf{x} \right), 
	\end{align}
	\begin{align}\label{property_two}
	H \left( z \big| E^{*} (\mathbf{x}), q_y (\cdot|E^{*} (\mathbf{x})) \right) = H \left( z \big|  q_y (\cdot|E^{*} (\mathbf{x})) \right),
	\end{align}
   \begin{align}\label{property_three}
	H \left( v \big| E^{*} (\mathbf{x}), q_y (\cdot|E^{*} (\mathbf{x})) \right) = H \left( v \big|  q_y (\cdot|E^{*} (\mathbf{x})) \right). 
	\end{align}
\end{proposition}
\begin{IEEEproof}
	When $ E $ is fixed, $ \mathcal{L}_{P^{*}} (E) = H \left( y|E(\mathbf{x}) \right) \ge H(y|\mathbf{x}) $, $ \mathcal{L}_{D^{*}} (E) = H \left( z \big| E(\mathbf{x}), q_y (\cdot|E(\mathbf{x})) \right) \le H \left( z \big| q_y (\cdot|E(\mathbf{x})) \right) $ and $ \mathcal{L}_{C^{*}} (E) = H \left( v \big| E(\mathbf{x}), q_y (\cdot|E(\mathbf{x})) \right) \le H \left( v \big| q_y (\cdot|E(\mathbf{x})) \right) $.
	Therefore, we obtain a lower bound of $ \mathcal{V}(E) $, that is
	\begin{align}
	\mathcal{V}(E) \ge  H(y|\mathbf{x})- \alpha H \left( z \big| q_y (\cdot|E(\mathbf{x})) \right) - \beta H \left( v \big| q_y (\cdot|E(\mathbf{x})) \right).
	\end{align}
	Since the lower bound is achieved if and only if all conditions \eqref{property_one}, \eqref{property_two} and \eqref{property_three} hold, proving that any optimal extractor $ E^{*} $ satisfies these conditions is identical to proving 
	\begin{align} \label{value function of best predictor}
	\mathcal{V}(E^{*}) = H(y|\mathbf{x})- \alpha H \left( z \big| q_y (\cdot|E^{*}(\mathbf{x})) \right) - \beta H \left( v \big| q_y (\cdot|E^{*}(\mathbf{x})) \right).
	\end{align}
	We note that the lower bound is achievable by considering a special case, where $ E^{*} (\mathbf{x}) = q_y (\cdot|\mathbf{x}) $, an extractor with the best representative ability over $ q_{train} (\mathbf{x}) $. In this case, we can check that the equality \eqref{value function of best predictor} holds.
\end{IEEEproof}

\begin{remark}
	Proposition 3 indicates that when all players are trained to be optimal and our adversarial model reaches equilibrium, the extractor $ E $ is able to extract all information about $ y $ from the training samples and eliminate any information about $ z $ and $ v $ except what is also related to $ y $.
\end{remark}

\begin{corollary} (Optimal outputs) If $ E $, $ P $, $ D $ and $ C $ have enough capacity and are trained to be optimal over $ q_{train} (\mathbf{x}) $, the outputs of our adversarial model achieve
	\begin{align}
	P_y (\cdot|E(\mathbf{x})) = q_y (\cdot|\mathbf{x}),
	\end{align}
	\begin{align}
	D_z \left( \cdot \big| E(\mathbf{x}), P_y (\cdot|E(\mathbf{x})) \right) = q_z (\cdot| q_y (\cdot|\mathbf{x})),
	\end{align}
	\begin{align}
	C_v \left( \cdot \big| E(\mathbf{x}), P_y (\cdot|E(\mathbf{x})) \right) = q_v (\cdot| q_y (\cdot|\mathbf{x})).
	\end{align}
\end{corollary}
\begin{IEEEproof}
	Based on Proposition 1, $ P\left( y|E(\mathbf{x})\right) =q(y|E(\mathbf{x})) $. According to Proposition 3, $ H \left( y| E (\mathbf{x}) \right) = H \left( y|\mathbf{x} \right) $, which implies that $ q(y|E(\mathbf{x})) = q(y|\mathbf{x}) $. Hence, $ P (y|E(\mathbf{x})) = q(y|\mathbf{x}) $.
	
	When both $ P $ and $ D $ are optimal, the equality \eqref{new loss of optimal D} holds, that is $ D \left( z \big| E(\mathbf{x}), P_y (\cdot|E(\mathbf{x})) \right) = q\left( z \big|E(\mathbf{x}), q_y (\cdot|E(\mathbf{x}))\right)  $. According to Proposition 3, the equality \eqref{property_two} holds, which is equivalent to $ q\left( z \big|E(\mathbf{x}), q_y (\cdot|E(\mathbf{x}))\right) = q(z|q_y (\cdot|E(\mathbf{x}))) $. Then, by considering $ P(y|E(\mathbf{x})) = q(y|\mathbf{x}) $, we achieve the equality $ D \left( z \big| E(\mathbf{x}), P_y (\cdot |E(\mathbf{x})) \right) = q(z|q_y (\cdot|\mathbf{x})) $. Similarly, we can have $ C \left( v \big| E(\mathbf{x}), P_y (\cdot|E(\mathbf{x})) \right) = q (v| q_y (\cdot|\mathbf{x})) $.
\end{IEEEproof}

\begin{algorithm} 
	\caption{Adversarial training for our multi-player neural network. } 
	\label{adversarial algorithm} 
	\begin{algorithmic}
		\STATE \textbf{Input:} Labeled samples $ \left\lbrace \left( \mathbf{x}_i,y_i,z_i,v_i \right)  \right\rbrace^{M}_{i=1} $, learning rates  $ \left( \eta_e,\eta_p,\eta_{d},\eta_c\right)  $ , hyperparameters $\left(  \alpha, \beta \right) $.
		\FOR{the number of training iterations}
		\STATE Sample a mini-batch of training data $ \left\lbrace \left( \mathbf{x}_i,y_i,z_i,v_i \right)  \right\rbrace^m_{i=1}  $
		\STATE Update the predictor $ P $:
		\STATE  \begin{center}
			$ \mathcal{L}^{i}_{P} \gets -\log P \left( y_i | E(\mathbf{x}_i) \right) $\\
			$ \Theta_p \gets \Theta_p - \eta_p \nabla_{\Theta_p} \frac{1}{m} \sum^{m}_{i=1} \mathcal{L}^{i}_{P} $
		\end{center}
		\FOR{the number of inner loops} 
		\STATE \begin{center}
			$ \mathbf{u}_i \gets P_y \left( \cdot | E(\mathbf{x}_i) \right) $ $ \blacktriangleright $ stop backpropagation\\
			$ \mathbf{o}_i = E(\mathbf{x}_i) \oplus \mathbf{u}_i $ $ \blacktriangleright $ concatenation
		\end{center}  
		\STATE Update the discriminator $ D $:
		\STATE \begin{center}
			$ \mathcal{L}^{i}_{D} \gets -\log D \left( z_{i}| \mathbf{o}_i \right)  $\\
			$ \Theta_{d}  \gets \Theta_{d} - \eta_{d} \nabla_{\Theta_{d}} \frac{1}{m} \sum^{m}_{i=1} \mathcal{L}^{i}_{D} $
		\end{center} 
		\STATE Update the discriminator $ C $:
		\STATE \begin{center}
			$ \mathcal{L}^{i}_{C} \gets -\log C \left( v_{i}| \mathbf{o}_i \right)  $\\
			$ \Theta_{c}  \gets \Theta_{c} - \eta_{c} \nabla_{\Theta_{c}} \frac{1}{m} \sum^{m}_{i=1} \mathcal{L}^{i}_{C} $
		\end{center} 
		\STATE Update the extractor $ E $:
		\STATE \begin{center}
			$ \mathcal{V}^{i} \gets  \mathcal{L}^{i}_{P} - \alpha \mathcal{L}^{i}_{D} - \beta \mathcal{L}^{i}_{C} $\\
			$ \Theta_e \gets \Theta_e - \eta_e \nabla_{\Theta_e} \frac{1}{m} \sum^{m}_{i=1} \mathcal{V}^{i} $
		\end{center}
		\ENDFOR
		\ENDFOR	
	\end{algorithmic} 
\end{algorithm}

\subsection{Adversarial Training Algorithm} 
Along the pipeline of theoretical analysis above, we design an adversarial training algorithm for our multi-player model. As depicted in Algorithm~\ref{adversarial algorithm}, it starts by updating the parameters of $ P $ in each training iteration and then optimizes those of $ D $, $ C $ and $ E $ in the inner loop. Since it is quite challenging to stabilize components in an adversarial model, especially the one with a minimax loss to optimize \cite{goodfellow2014generative,nguyen2017dual}, we have $ E $ in the inner loop for extra training to refine it. Furthermore, the model could minimize the value function $ \mathcal{V} $ by increasing the losses $ \mathcal{L}_{D} $ and $ \mathcal{L}_{C} $ improperly, which is known as model collapse phenomenon. To avoid this situation, we also update $ D $ and $ C $ in the inner loop, which aims at leading $ \mathcal{L}_{D} $ and $ \mathcal{L}_{C} $ to descend in right directions. Additionally, at the beginning of each inner iteration, an intermediate variable $ \mathbf{u}_i $ is assigned to be $ P_y (\cdot | E(\mathbf{x}_i)) $ and thereafter concatenated with $ E(\mathbf{x}_i) $ as input $ \mathbf{o}_i $ for two adversaries. These operations can effectively detach $ P $ from $ D $ and $ C $, and thus prevents gradients from propagating back to $ P $ as aforementioned. 

In our experiment, we build our multi-player network and implement the adversarial training algorithm using Python with PyTorch packages. The training data consists of about five thousand samples and is transformed with the z-score normalization before training. We empirically set the mini-batch size to 750, each hyperparameter to 0.5 and each learning rate to 0.001. Finally, we update the parameters $ \Theta_e $, $ \Theta_d $ and $ \Theta_c $ with more than five thousand iterations.

%use Adam optimizers with weight decay of 0.01 to

\section{Evaluation in Real Environments}\label{sec: evaluation}

%In this section, we empirically evaluate our adversarial network based device authentication system on real-world radio traces. In our experiments, radio measurements are recorded under different static and dynamic body motions in various indoor and outdoor environments. 

\subsection{Experimental Methodology}
\textbf{Implementation.} We implement a proof-of-concept prototype of the proposed system with three GNURadio/USRP B210 devices. These devices are configured to communicate in the 2.4 GHz ISM band with a sampling rate of 500 Hz, which is feasible for most commercial wearable devices. Moreover, two USRP devices are placed on a volunteer, referred to as a legitimate user. Specifically, one of them locates at the left side of the user's waist as an on-body receiver, and the other device is carried by the user's right hand as an on-body transmitter. The remaining USRP device is held by another volunteer, referred to as a malicious user, and it is regarded as an off-body transmitter.

\textbf{Data collection.} We collect radio traces under different surrounding environments and user body motions. The experimental environments encompass three indoor settings, a \textit{laboratory}, an \textit{office} and a \textit{corridor}, and two outdoor ones, a \textit{rooftop} and a \textit{park}. In each environment, the legitimate user, carrying the on-body transceivers, is asked to perform controlled and uncontrolled motions, respectively. In the controlled scenario, the legitimate user is confined to take five basic motions, including two static actions, \textit{sitting} and \textit{standing}, and three dynamic ones, \textit{arm moving}, \textit{rotating} and \textit{walking}. In the uncontrolled scenario, the legitimate user can impose whatever body motions he or she likes. However, the malicious user, holding the off-body transmitter, is not restricted to any specified motion throughout the experiment and can walk freely in the proximity of 1-5 meters away from the legitimate user. When collecting data, we ask two users to take their own motions for one minute, during which we control one of the on- and off-body transmitters to broadcast signals and use the receiver to record corresponding radio traces. The above trial is repeated for 20 times in each motion setting, and the participants are given a rest period of around 30s between two consecutive trials. Finally, we conduct our experiment over seven days with five volunteers, including two females and three males, and yield radio traces of ten hours in total.

\textbf{Dataset.} We partition the collected on- and off-body traces into RSS segments and extract propagation profiles from these segments according to Section~\ref{sec: profile}. Then, we label the extracted on- and off-body profiles with respect to corresponding motions and environments and obtain a total of 7200 labeled samples for our adversarial network. Therein, 6000 samples are from the controlled user motion scenario, and 1200 samples are from the uncontrolled scenario. When training and testing our model, we randomly take out 4800 samples from the controlled scenario for training and combine the leftover 1200 ones and all 1200 samples from the uncontrolled scenario for testing. Moreover, the numbers of on- and off-body samples are equal in both the training and testing sets.
 
\textbf{Evaluation metrics.} To demonstrate the performance of the proposed system, we use \textit{accuracy}, \textit{true positive (TP) rate} and \textit{false positive (FP) rate} as metrics, which are given as below. 
\begin{itemize}
	\item \textbf{Accuracy.} Accuracy is defined as the ratio of the number of correctly classified RSS segments to the total number of on- and off-body segments.
	\item \textbf{TP rate.}  TP rate denotes to the ratio of the number of correctly detected on-body RSS segments to the total number of on-body segments.
	\item \textbf{FP rate.} FP rate is computed as the ratio of the number of mistakenly recognized off-body RSS segments to the total number of off-body segments.
\end{itemize}
 
\subsection{Performance Results}

\begin{figure}
	\centering
	\includegraphics[width=0.5\linewidth]{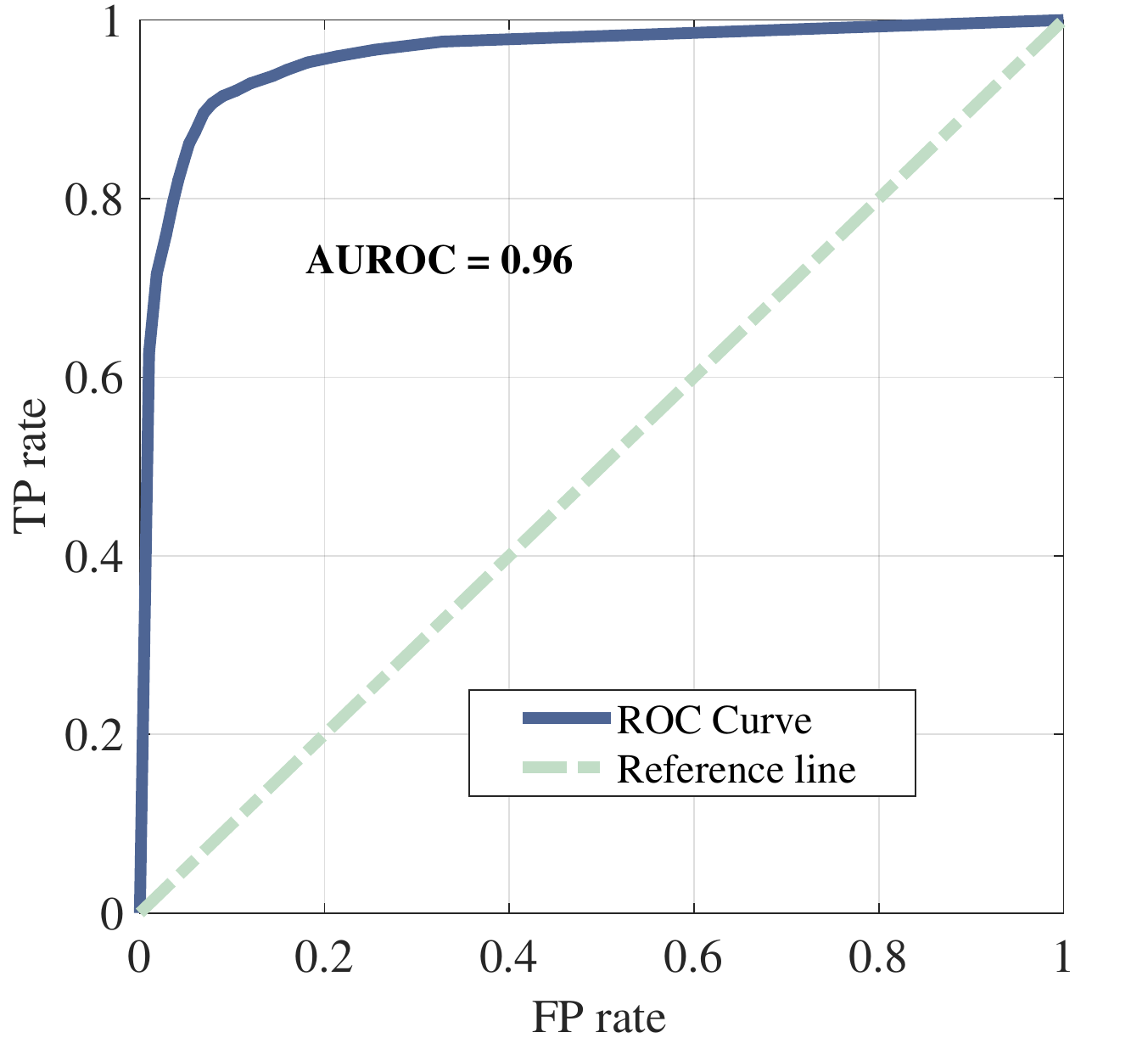}
	\caption{ROC curve of the proposed system. The corresponding AUROC is 0.96 and the reference line stands for the random guessing model.}
	\label{fig:roc}
\end{figure}

\begin{table}[h]
	\centering
	\small
	\caption{Overall Performance Results}\label{tab:overall_results}
	\begin{tabular}{c|c|c}
		\hline
		\textbf{Accuracy} & \textbf{TP Rate} & \textbf{FP Rate} \\
		\hline
		91.6\% $ \pm $ 2.4\%  & 90.6\% $ \pm $ 1.9\%  &  7.2\% $ \pm $ 2.8\%\\
		\hline
	\end{tabular}
\end{table}

We train our model on the collected training dataset and run the trained model on the testing dataset to obtain prediction results of all testing samples. Specifically, the training dataset only consists of samples from the five controlled motions in all environments. Besides controlled samples, the testing dataset contains samples from the uncontrolled scenarios, which are never used to train our model.

\textbf{Overall performance.} We first illustrate the overall performance of our authentication system. Specifically, based on all prediction results and their environment and motion labels, we can average accuracies, TP and FP rates in all motion-environment scenarios and obtain the results in Table~\ref{tab:overall_results}. As shown in Table~\ref{tab:overall_results}, our system is able to identify 91.6\% of on- and off-body devices on average. Specifically, it can correctly recognize on-body devices with a ratio of 90.6\% and successfully mitigate 92.8\% of attack attempts from off-body devices. Since on-body authentication is a binary classification task, we further use the receiver operating characteristic (ROC) curve to measure how well our system can correctly discriminate on- and off-body samples from different scenarios. The ROC curve depicts TP rates against FP rates at various threshold settings and tells a classifier's capability of distinguishing between two classes. For a good classifier, high TP and low FP rates are expected when the threshold is in $ \left(0,1 \right)  $. As depicted in Fig.~\ref{fig:roc}, our system's ROC curve first goes straight up and then becomes steady promptly as the FP rate increases. Besides, the area under the ROC curve (AUROC) reaches 0.96, which is close to 1, i.e., the AUROC of the ideal classifier. The above results indicate that our system achieves a good discrimination ability for on- and off-body samples under different motions and environments.

\begin{figure}%[t] 
	\centering
	\subfigure[Accuracy for different body motions.]{
		\includegraphics[width=0.45\linewidth]{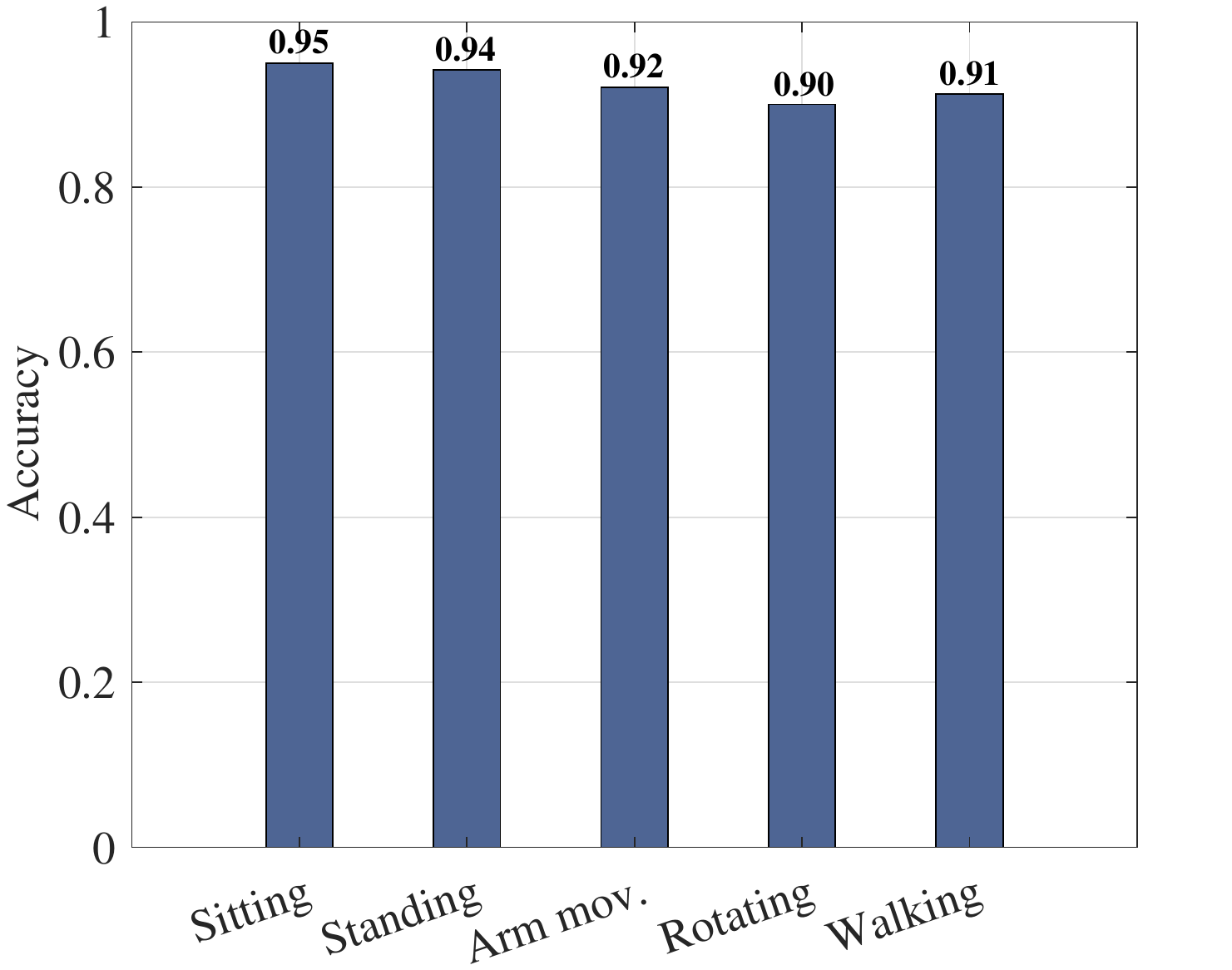}}
	\label{4a}
	%\hspace{0.3cm}
	\subfigure[TP and FP rates for different body motions.]{
		\includegraphics[width=0.46\linewidth]{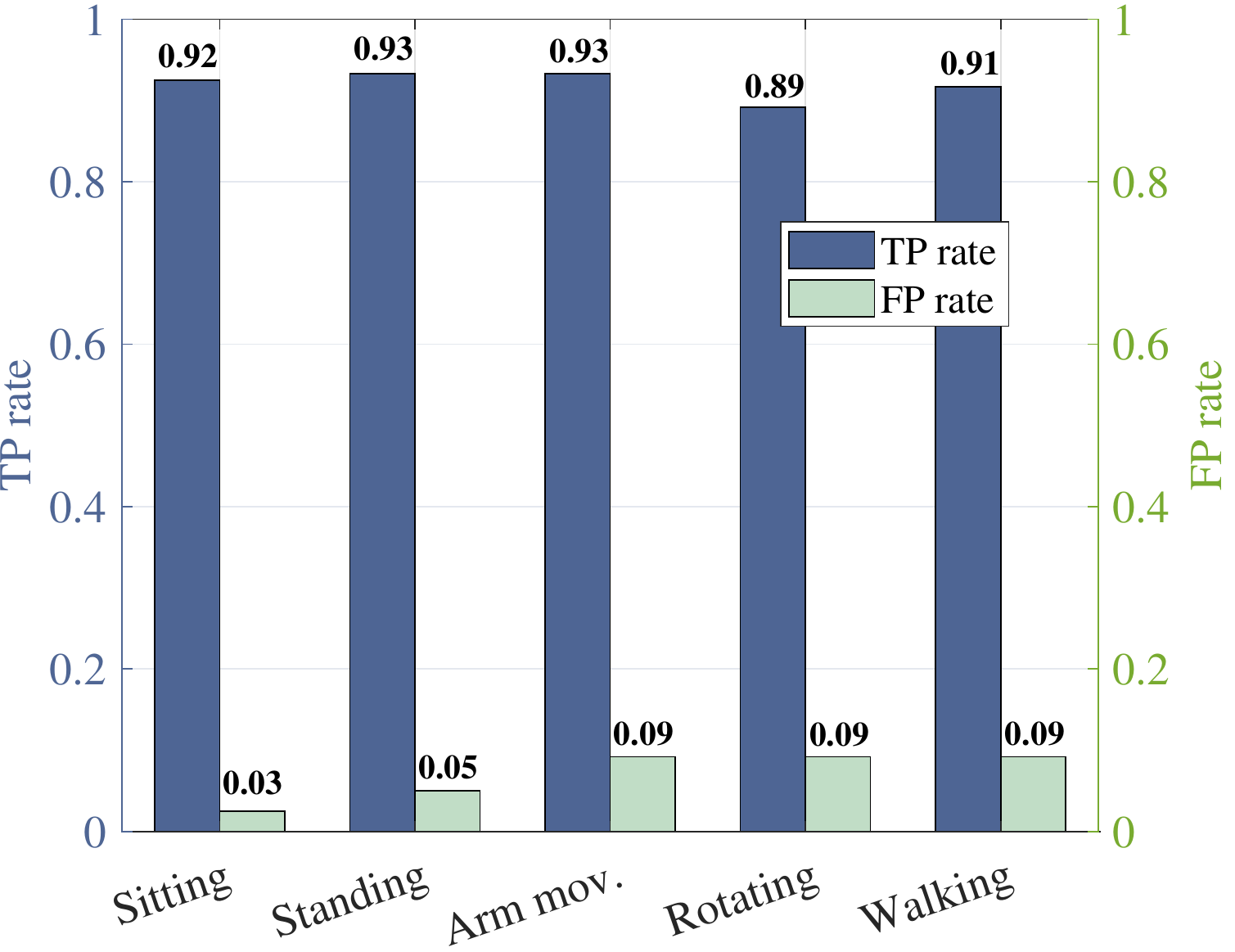}}
	\label{4b}
	\caption{Performance under different body motions.}
	\label{fig:Performance for motions} 
\end{figure}

\textbf{Performance under different motions.} We then elaborate on the system's performance under each frequently appearing motion. In general, each selected motion has a unique movement pattern of the human body and thus exhibits a different effect on BAN radio waves. We divide all prediction results into different motion groups based on the motion label and calculate the evaluation metrics in each group. As plotted in Fig.~\ref{fig:Performance for motions} (a), we observe that the proposed system achieves better performance for the static motions than for the dynamic ones. The same observations can be found in Fig.~\ref{fig:Performance for motions} (b). Therein, higher TP and lower FP rates are clearly present in the static states. It is due to that body motions have a great impact on the attenuation of on-body propagations as explained before, and there are fewer disturbances caused by body movements in radio signals when the user sits or stands still with IoT devices, which makes it much easier for the proposed system to recognize on- and off-body propagation patterns. Despite the above differences, the system still achieves average TP and FP rates of 92.0\% and 6.0\%, respectively, in the controlled user motion scenario.

\begin{figure}
	\centering
	\includegraphics[width=0.43\linewidth]{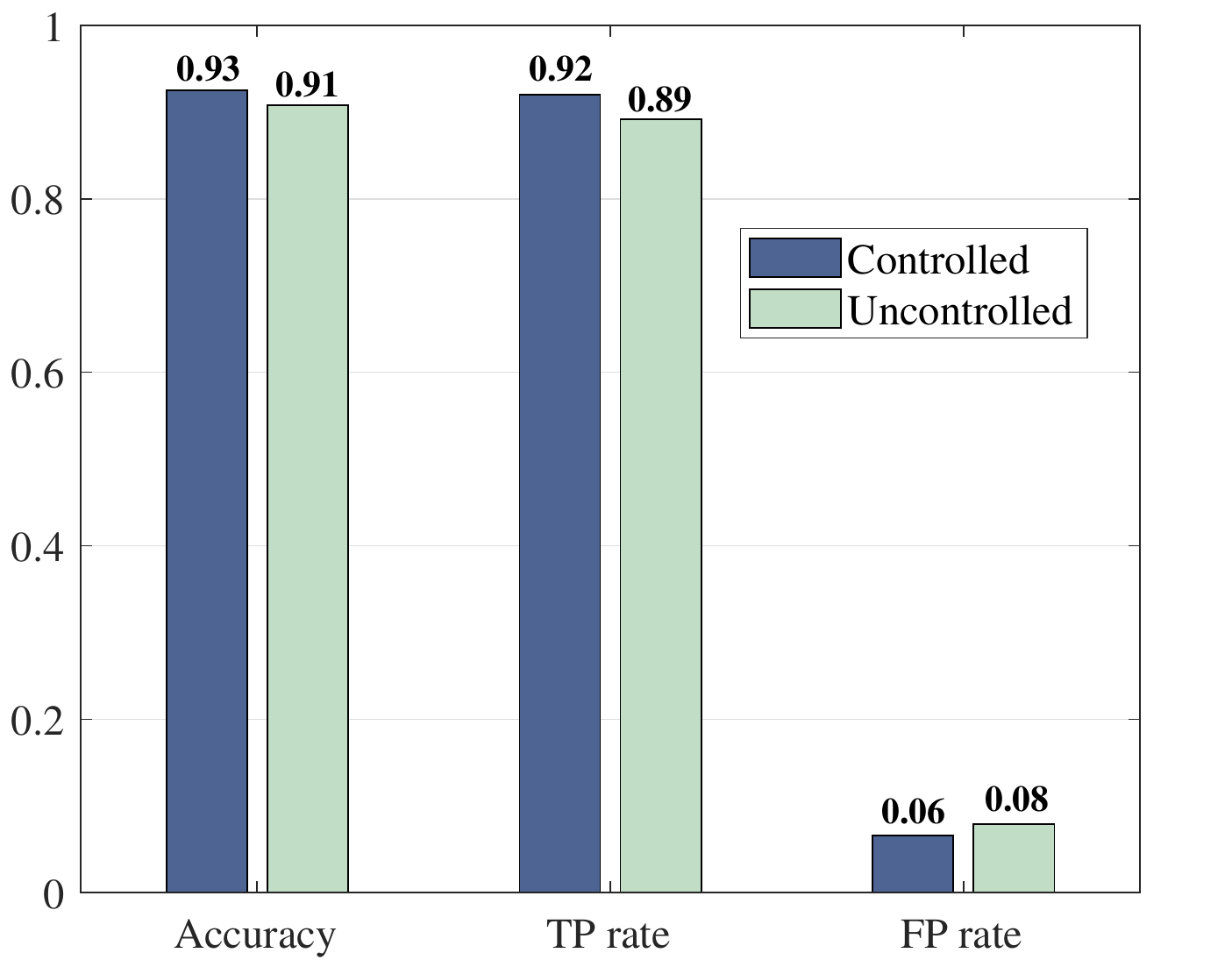}
	\caption{Performance in the controlled and uncontrolled motion scenarios.}
	\label{fig:uncontrolled performance}
\end{figure}

Next, we compare the system performance in the uncontrolled user motion scenario with that in the controlled scenario. As illustrated in Fig.~\ref{fig:uncontrolled performance}, the system shows performance degradation in terms of accuracy, TP and FP rates in the uncontrolled scenario. The reason for the performance degradation is that more irregular and complicated body movements are present when the user behaves casually with IoT devices, which makes the feature extractor to extract more noisy features about radio propagation patterns and thus hampers the prediction ability of the on-off predictor. More specifically, the system has a TP rate reduction of 3.0\% and a FP rate increase of 2.0\% for uncontrolled motions. This is due to the fact that, compared with off-body radio signals, on-body signals, dominated by creeping waves, are more sensitive to user motion dynamics, which results in more on-body RSS segments to be mistakenly classified as off-body ones. 

\begin{figure}%[t] 
	\centering
	\subfigure[Accuracy for different environments.]{
		\includegraphics[width=0.43\linewidth]{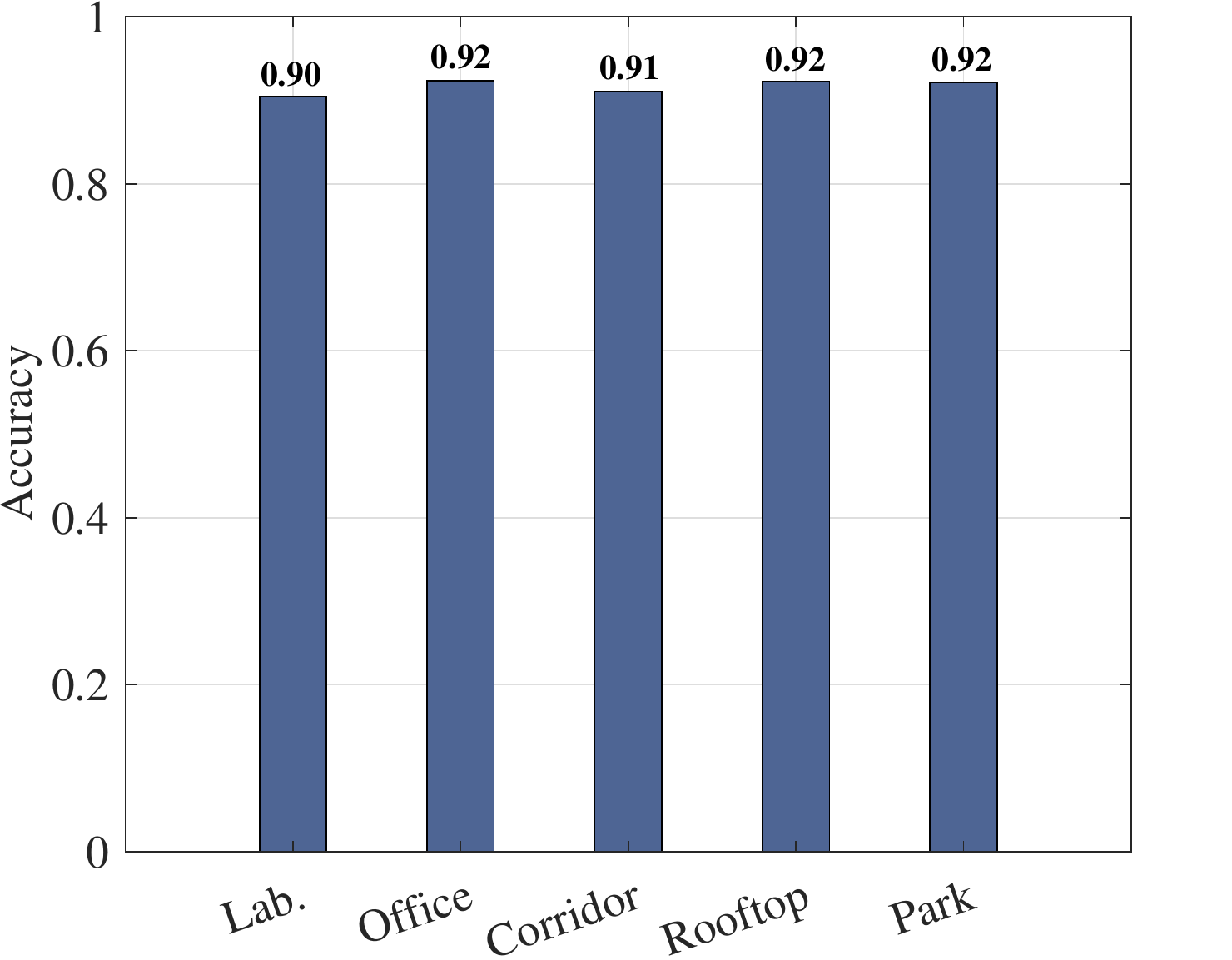}}
	\label{5a}
	\hspace{0.3cm}
	\subfigure[TP and FP rates for different environments.]{
		\includegraphics[width=0.44\linewidth]{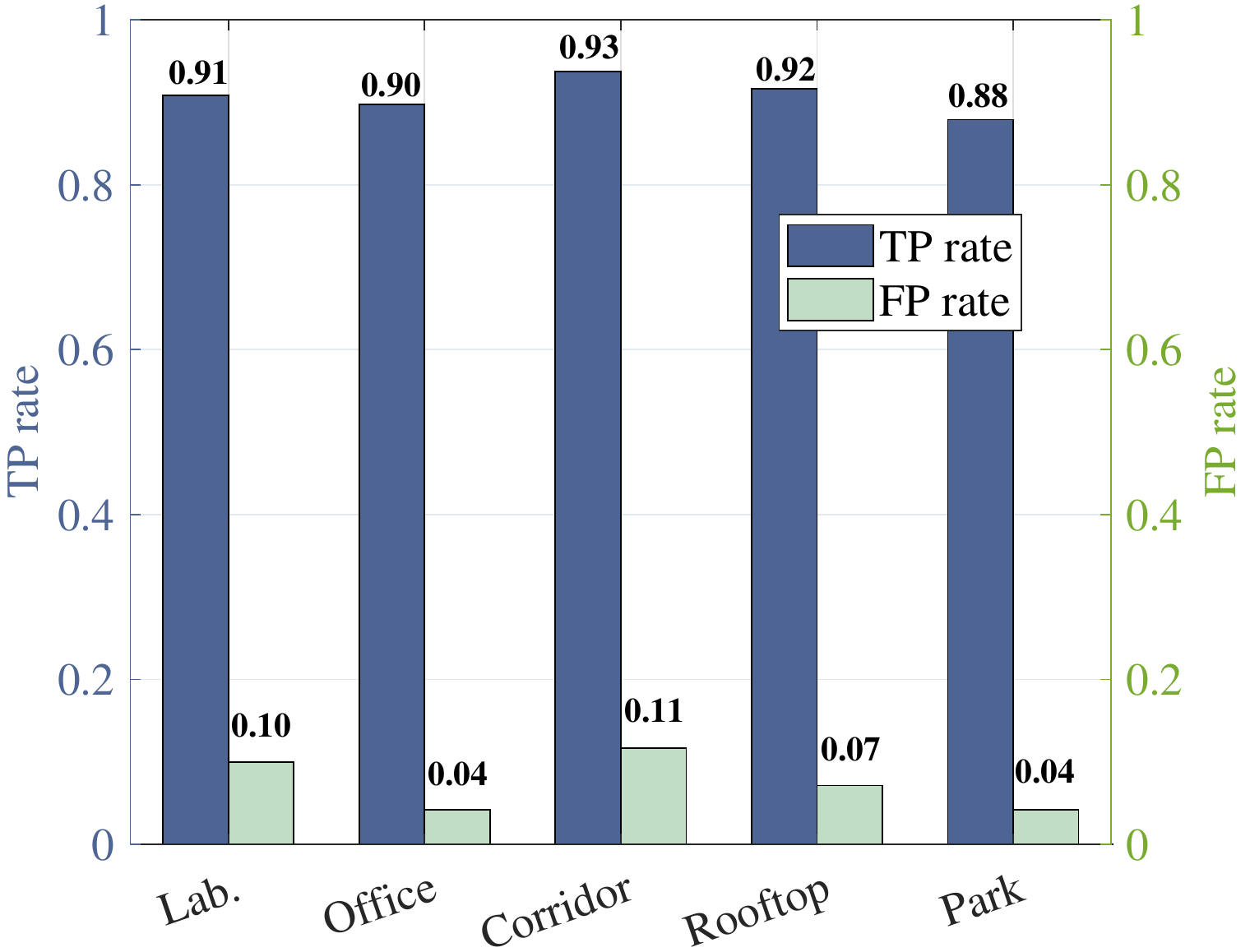}}
	\label{5b}
	\caption{Performance in different environments.}
	\label{fig:Performance for environments} 
\end{figure}

\textbf{Performance in various environments.} We next validate the authentication performance of our system in various indoor and outdoor environments in Fig.~\ref{fig:Performance for environments}. Basically, indoor radio propagations remarkably differ from outdoor propagations in terms of shadowing and multipath fading. We divide the prediction results into different environment groups based on the environment label and compute the evaluation metrics in each group. As shown in Fig.~\ref{fig:Performance for environments} (a), our system achieves almost the same accuracy in each environment. However, as illustrated in Fig.~\ref{fig:Performance for environments} (b), the differences between the indoor and outdoor environments can be reflected more clearly on FP rates. Roughly speaking, lower FP rates are found in the outdoor settings rather than in the indoor ones. This is because that there generally exists less multi-path variations in outdoor propagations, which consequentially leads to less off-body segments to be mistakenly recognized as on-body ones. However, according to Fig.~\ref{fig:Performance for environments} (b), the office setting shows the lower FP rate than those of the other indoor environments due to less disturbances caused by other people. The above observations indicate that off-body segments tend to be more susceptible to environmental dynamics. Moreover, environment noise in RSS segments is an important factor that influences the system's performance. Generally, the lower the signal-to-noise ratio (SNR) is, the less distinguishability on- and off-body RSS segments have and the more difficult it is for a classifier to discriminate. To deal with environment noise, our system extracts representative time and frequency domain features from noisy RSS segments and further uses an environment discriminator to exclude environment specific information. As shown in Fig.~\ref{fig:Performance for environments} (a), our system achieves an accuracy of over 90\% in each environment, which shows the effectiveness of our system in the presence of real-world environment noise.

\begin{figure}%[t] 
	\centering
	\subfigure[Training losses of discriminators.]{
		\includegraphics[width=0.43\linewidth]{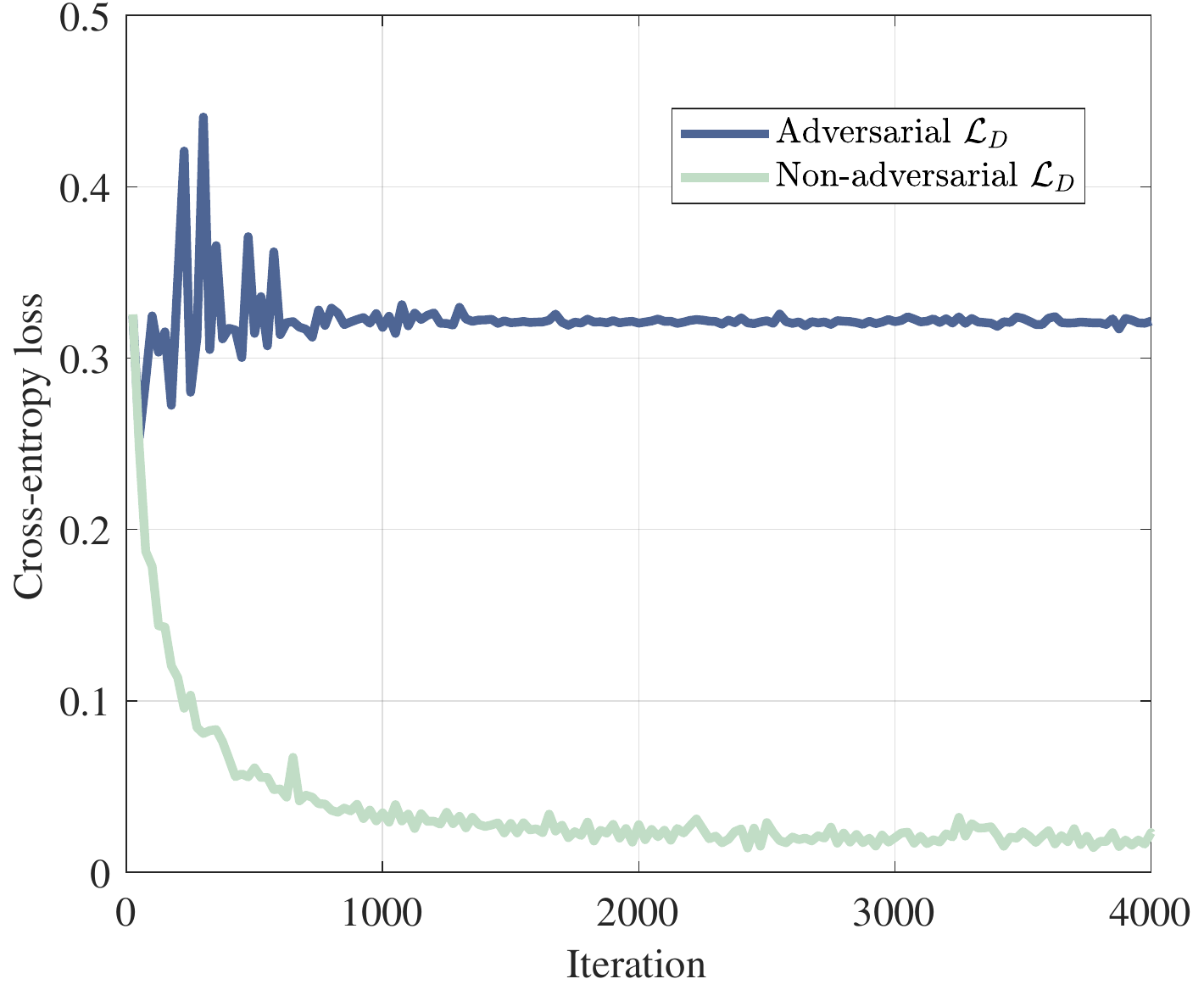}}
	\label{6a}
	%\hspace{0.3cm}
	\subfigure[Training losses of classifiers.]{
		\includegraphics[width=0.43\linewidth]{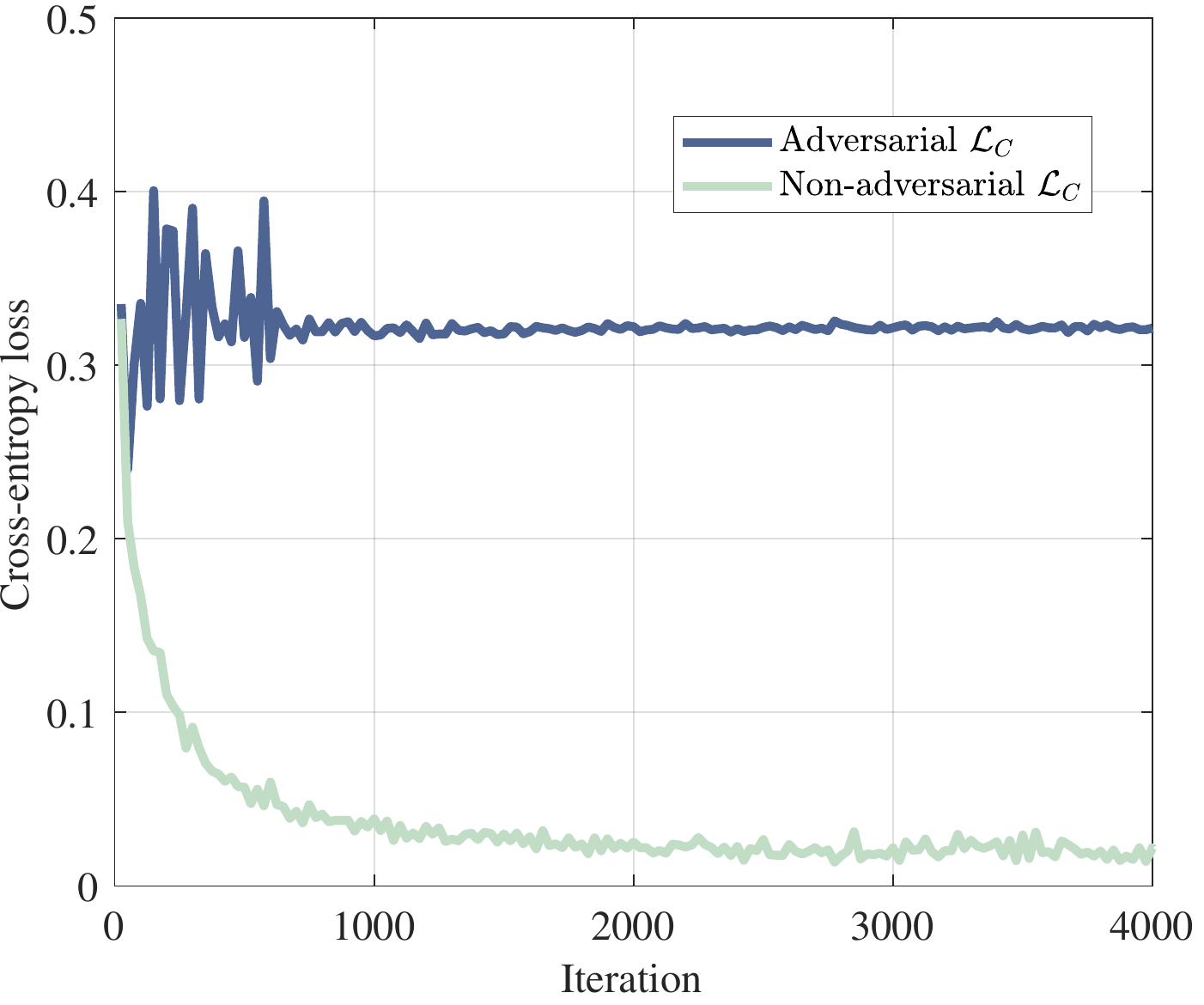}}
	\label{6b}
	\caption{Training losses of adversarial and non-adversarial models.}
	\label{fig:training and testing losses} 
\end{figure}

\begin{figure}
	\centering
	\includegraphics[width=0.43\linewidth]{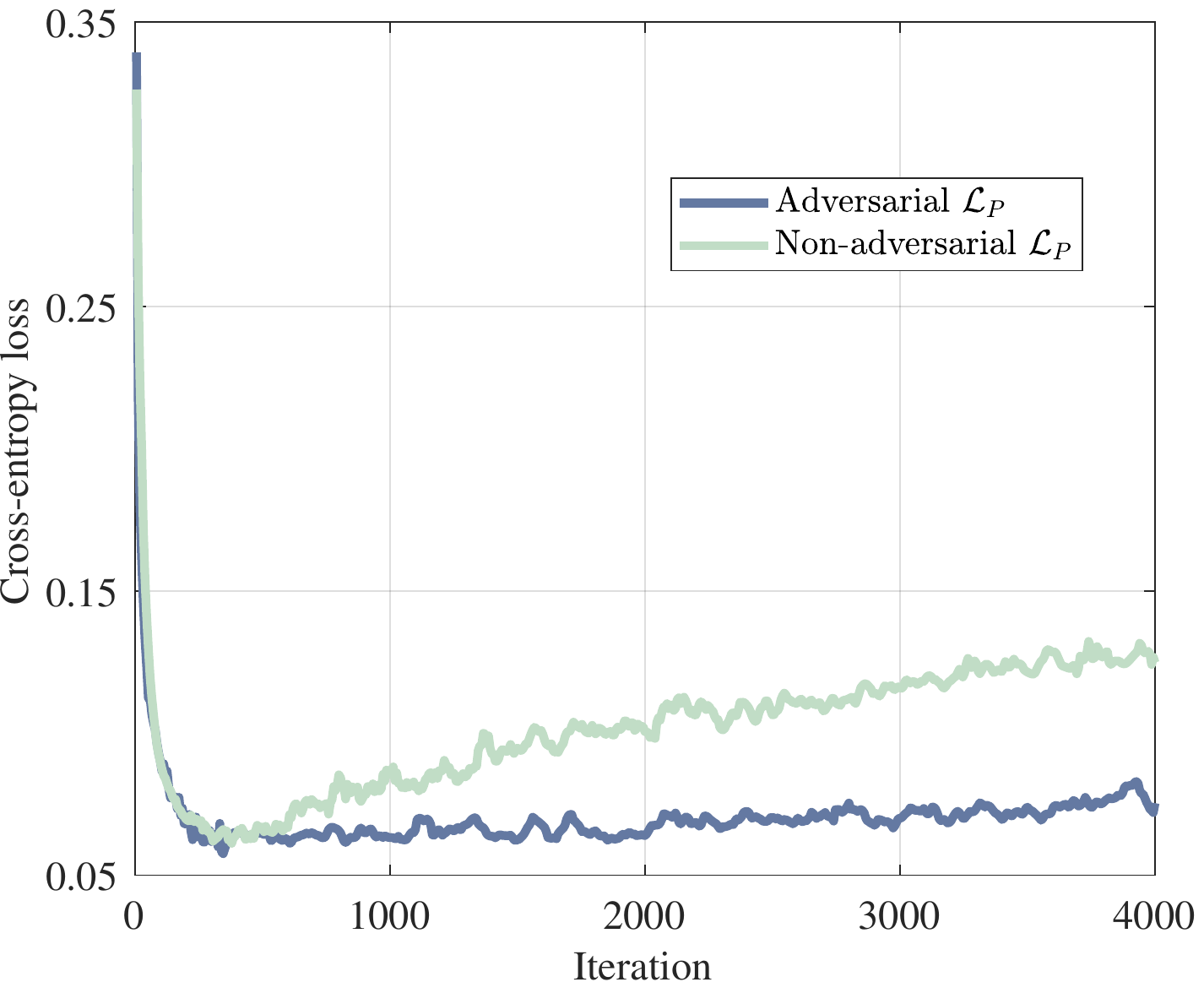}
	\caption{Testing losses of predictors. The rise of curves is due to the over-fitting phenomenon.}
	\label{fig:loss predictor}
\end{figure}

\textbf{Effectiveness of minimax game.} We further illustrate the benefits of adopting two adversaries in our multi-player model. Our adversarial discriminator and classifier aim at helping the feature extractor to discover transferable features and boosting the generalization ability of the on-off predictor. To illustrate these merits, we set up a version of our model with a pair of non-adversarial discriminator and classifier as a baseline. Note that in the baseline model, the update of the extractor' parameters relies solely on the minimization of the predictor's loss.  

Fig.~\ref{fig:training and testing losses} (a) and (b) plot the training losses of discriminators and classifiers in our and baseline models, respectively. In each figure, the lower the value is, the more the information pertaining to body motions or ambient environments is learned. In Fig.~\ref{fig:training and testing losses} (a), we can see that the loss of the non-adversarial discriminator declines quickly and then stabilizes at a very low level. However, the loss of our adversarial discriminator first fluctuates dramatically and then finally converges to a high value. The same observations can be found in Fig.~\ref{fig:training and testing losses} (b). This is due to the fact that at the beginning of training process, the fluctuations of an adversarial loss are incurred by its minimax optimization, and they mitigate gradually as motion or environment specific features irrelevant to the predictor fade out in the extracted feature representation. The above results reveal that the extractor in our model abstracts more transferable features than that in the baseline. Furthermore, comparing the performance of two predictors in our and baseline models in Fig.~\ref{fig:loss predictor}, we find that both loss curves decrease at first and then increase after certain numbers of iterations. However, the adversarial curve rises up at a slower speed than the non-adversarial one, which suggests that the adversarial discriminator and classifier work as two regularizers for alleviating over-fitting and enable the promotion of the predictor's generalization ability.

\section{Conclusion}\label{sec: conclusion}
This paper proposes a new device authentication system that takes one step forward to embrace the advent of human-centric IoT by supporting various wearable devices anytime and anywhere. The key enabling technique is using an adversarial multi-player network to effectively recognize radio propagation patterns under diverse user motions in different environments. Moreover, integrating with upper-layer security protocols, our system is able to in depth secure on-body IoT device pairing and data transmission. We theoretically analyze our adversarial model and prove that at equilibrium, our model becomes invariant to motion variances and environment changes. We build a working prototype of our system using USRP devices and conduct extensive experiments with various static and dynamic user motions in typical indoor and outdoor settings. The experimental results show that our system can successfully identify 90.6\% of legitimate devices and mitigate 92.8\% of active attack attempts.

\bibliographystyle{IEEEtran}
\bibliography{IEEEabrv,./Onbody_authentication}

\end{document}